\documentclass[lettersize,journal]{IEEEtran}

\usepackage{url}
\usepackage{graphicx,subfigure}
\usepackage{epstopdf}
\usepackage{algpseudocode}
\usepackage{amsmath}
\usepackage{cite}
\def\BibTeX{{\rm B\kern-.05em{\sc i\kern-.025em b}\kern-.08em
    T\kern-.1667em\lower.7ex\hbox{E}\kern-.125emX}}

\usepackage{amssymb}
\usepackage{amsthm}
\usepackage[caption=false,font=normalsize,labelfont=sf,textfont=sf]{subfig}
\usepackage{booktabs}
\usepackage[table]{xcolor}
\usepackage{enumitem}

\usepackage{amsfonts}
\usepackage{multirow}
\usepackage[normalem]{ulem}
\usepackage{color}
\usepackage{array}
\usepackage{acro}
\usepackage{soul}
\usepackage[algo2e]{algorithm2e} 

\SetKwInput{KwData}{Data}

\usepackage[underline=true]{pgf-umlsd}

\usepackage{setspace}

\DeclareRobustCommand{\acrodef}[2]{\DeclareAcronym{#1}{short=#1,long=#2}}
\acrodef{RRC}{Radio Resource Control}
\acrodef{ZMQ}{ZeroMQ}
\acrodef{LSTM}{Long Short-Term Memory}
\acrodef{LAL}{Listen-and-Learn}
\acrodef{TS}{Technical Specifications}
\acrodef{NR}{New Radio}
\acrodef{eMBB}{enhancing Mobile Broadband}
\acrodef{mMTC}{massive Machine Type Communications}
\acrodef{O-RAN}{Open Radio Access Network}
\acrodef{NAS}{Non-Access Stratum}
\acrodef{AS}{Access Stratum}
\acrodef{UE}{User Equipment}
\acrodef{MITM}{man-in-the-middle}
\acrodef{OTA}{over-the-air}
\acrodef{CN}{core network}
\acrodef{gNB}{gNodeB}
\acrodef{DoS}{Deny of Service}
\acrodef{BS}{Base Station}
\acrodef{NSA}{Non Standard-Alone}
\acrodef{SA}{Standard-Alone}
\acrodef{EPS}{Evolved Packet System}
\acrodef{eNB}{evolved NodeBs}
\acrodef{AKA}{authentication and key agreement}
\acrodef{EC-AKA}{Ensured confidentiality Authentication and Key agreement}
\acrodef{ASME}{Access Security Management Entity}
\acrodef{UP}{User Plane}
\acrodef{KDF}{key derivation function}
\acrodef{IMSI}{international mobile subscriber identity}
\acrodef{S-TMSI}{SAE Temporary Mobile Subscriber Identity}

\usepackage{algorithm,algcompatible}
\algnewcommand\algorithmicreturn{\textbf{return}}
\algnewcommand\RETURN{\algorithmicreturn}
\algnewcommand\algorithmicprocedure{\textbf{procedure}}
\algnewcommand\PROCEDURE{\item[\algorithmicprocedure]}%
\algnewcommand\algorithmicendprocedure{\textbf{end procedure}}
\algnewcommand\ENDPROCEDURE{\item[\algorithmicendprocedure]}%
\algnewcommand{\algvar}[1]{{\text{\ttfamily\detokenize{#1}}}}
\algnewcommand{\algarg}[1]{{\text{\ttfamily\itshape\detokenize{#1}}}}
\algnewcommand{\algproc}[1]{{\text{\ttfamily\detokenize{#1}}}}
\algnewcommand{\algassign}{\leftarrow}
\MakeRobust{\Call}

\begin{document}


\title{Formal-Guided Fuzz Testing: Targeting Security Assurance from Specification to Implementation for 5G and Beyond}


\author{Jingda Yang, {\it Student Member, IEEE}, Sudhanshu Arya, {\it Member, IEEE}, and Ying Wang, {\it Member, IEEE}
\thanks{J. Yang, S. Arya, and Y. Wang are with the School of Systems and Enterprises, Stevens Institute of Technology, Hoboken, NJ, USA 07030. The corresponding author is Y. Wang (ywang6@stevens.edu).  This effort was sponsored by the Defense Advanced Research Project Agency (DARPA) under grant no. D22AP00144.}
}


\maketitle
\begin{abstract}
Softwarization and virtualization in 5G and beyond necessitate thorough testing to ensure the security of critical infrastructure and networks, requiring the identification of vulnerabilities and unintended emergent behaviors from protocol designs to their software stack implementation. Formal methods demonstrate efficiency in abstracting specification models at the protocol level, while fuzz testing provides comprehensive experimental evaluations of system implementations. However, the state of art formal and fuzz testing are both labor-intensive or computationally complex. To provide an efficient and comprehensive solution, we propose a novel and first-of-its-kind approach that connects the strengths and coverage of formal and fuzzing methods to efficiently detect vulnerabilities across protocol logic and implementation stacks in a hierarchical manner. We design and implement formal verification to detect attack traces in critical protocols, which are used to guide subsequent fuzz testing and incorporate feedback from fuzz testing to broaden the scope of formal verification. This innovative approach significantly improves efficiency and enables the auto-discovery of vulnerabilities and unintended emergent behaviors from the 3GPP protocols to software stacks. In this paper, we demonstrate this approach to the 5G Non-Stand-Alone (NSA) security processes, which have more complicated designs and higher risks due to the compatibility requirement to legacy and exiting 4G networks compared to 5G Stand-Alone (SA) processes, with a focus on the Radio Resource Control (RRC), Non-access Stratum (NAS) and Access Stratum (AS)  authentication process. Through formal analysis, we identify protocol-level vulnerabilities related to user credentials disclosure and man-in-the-middle (MITM) attack. Subsequently, we employ bit-level fuzzing to assess the potential impacts and risks of identifier variation susceptible to integrity vulnerabilities, followed by command-level mutation-based fuzzing, assuming fixed identifier values, to evaluate the potential impacts and risks associated with confidentiality vulnerabilities. 
Following this approach, we discover one identifier leakage model, one DoS attack model, and two eavesdrop attack models due to the absence of rudimentary MITM protection within the protocol, despite the existence of a Transport Layer Security (TLS) solution to this issue for over a decade. More  remarkably, guided by the identified formal analysis and attack models, we exploit $61$ vulnerabilities using fuzz testing demonstrated on srsRAN platforms. These identified vulnerabilities contribute to fortifying protocol-level assumptions and refining the search space. Compared to state-of-the-art fuzz testing, our united formal and fuzzing methodology enables auto-assurance by systematically discovering vulnerabilities. It significantly reduces computational complexity, transforming the non-practical exponential growth in computational cost into linear growth. Our formal-guided fuzz testing system provides a robust and self-reinforcing solution to the scalability challenges that often arise when detecting vulnerabilities and unintended emergent behaviors in intricate, large-scale 5G systems and their deployments in critical infrastructures and verticals.  
\end{abstract}

\section{Introduction}

 \IEEEPARstart{V}{erticals} in 5G and next-generation infrastructure create a diverse and intricate environment consisting of software, hardware, configurations, instruments, data, users, and various stakeholders~\cite{Alcaraz-Calero2018Leading5G-PPP}. With system complexity and its lack of security emphasis by domain scientists, the formed ecosystem requires a comprehensive evaluation and validation for improved research and transitional Critical Infrastructure (CI) security posture~\cite{shatnawi2022digital}. 

Despite two major state-of-the-art approaches, formal verification and fuzz testing, being proposed to detect various vulnerabilities and unintended emergent behaviors of the 5G network, limitations in large-scale systems and stacks still exist. Formal verification can provide a high-level concept of protocol and logical proof of security and vulnerability~\cite{Hussain20195Greasoner:Protocol}. In contrast, fuzz testing can offer a detailed and comprehensive experimental platform, detecting potential vulnerabilities in the 5G code implementation platform~\cite{klees2018evaluating}. However, open issues and challenges of pick-and-choose fuzz testing and formal analysis in various scenarios still exist~\cite{souri2019state,beaman2022fuzzing}.

By unifying the fuzz testing with the formal analysis, it becomes possible to initiate a reciprocal cycle between the two approaches, leading to the identification of vulnerabilities across the entire search space. Formal verification can offer valuable guidance and assumptions to fuzz testing, while fuzz testing can broaden the scope of formal verification. The unification should be complementary and enable mutual amplification. As a result, vulnerability and unintended emergent behavior detection could be extended with scalability when amplification occurs. With an objective to improve the scalability, we propose an innovative heuristic approach by integrating fuzz testing with formal analysis. The proposed technique overcomes the limitations of the fuzz testing and the formal analysis and thereby enables the model checkers to detect a wide range of vulnerabilities in large complex 5G systems.
    
   
    
    

In the subsequent sections of this paper, we provide a concise overview of the structure of our proposed comprehensive formal verification and fuzz testing integrated vulnerability detection framework (Section~\ref{system_desin}). Subsequently, we elucidate the mechanism behind our proposed dependency-based protocol abstraction and evaluation approach (Section~\ref{protocol_abstraction}), followed by presenting examples of dependency analysis (Section~\ref{dependency_analysis}). Furthermore, we apply the dependency-based protocol abstraction and evaluation approach to the \ac{NSA} 5G communication establishment process (Section~\ref{formal_analysis}), where we present and analyze the results of formal verification (Section~\ref{attack_model}). Additionally, we propose proven or novel solutions for each detected formal attack model. Subsequently, leveraging the identified assumptions, we apply our proposed fuzz testing framework to verify and analyze the implementation of the \ac{NSA} 5G communication establishment process (Section~\ref{fuzz_test}). Lastly, in Section~\ref{comparsion}, we utilize intuitive visualizations to analyze the efficiency of different fuzzing strategies across various fuzzing scopes.

\section{Related Work and Background}\label{Background}
5G technologies are of rapidly increasing importance to the national and regional infrastructure and offer unprecedented connectivity benefits. However, these technologies also present an attack surface of unprecedented size due to the complexity of both specifications and implementations of 5G stacks. Previous researchers proposed vulnerability detection approaches\cite{Hussain20195Greasoner:Protocol,Wang2021AI-PoweredOptimization,Wang2022AnonymousShetty}, among which two categories have been intensively researched: formal verification and fuzz testing. 

Formal verification is the technology that transfers natural language-defined protocols into symbolic logic language, which is feasible to establish the validity of the given proposition through a finite process of mathematical verification. Several formal analysis frameworks in the existing research are proposed to determine which security guarantees are satisfied in 5G protocols by applying formal methods and automated verification in the symbolic model, like Tamarin\cite{Meier2013TheProtocols}, and 5G reasoner\cite{Cremers2019Component-BasedConfusion}. Hussian\cite{Hussain20195Greasoner:Protocol} et al. even proposed a cross-layer formal verification framework, which combines model checkers and cryptographic protocol verifies through the application of the abstraction-refinement principle. Besides formal verification frameworks, different formal strategies are introduced to prove the security assumption set, like~\cite{peltonen2021comprehensive}. For example, the pre-authentication message sent unencrypted has been acknowledged as the root cause of many known LTE and 5G protocol exploits\cite{Labib2017EnhancingProcess,Rupprecht2019BreakingTwo,Shaik2017PracticalSystems}. Furthermore, Some registration and access control protocols, including \ac{AKA}, RRC, etc. have been applied formal methods in various framework\cite{Basin2018AAuthentication,Cremers2019Component-BasedConfusion,Hussain20195Greasoner:Protocol}. When applied in the 5G security design, necessary lemmas are verified helping lemmas, sanity-check lemmas, and the lemmas that check the relevant security properties against the 5G protocols\cite{Basin2018AAuthentication}. 

A Fuzz tester (or fuzzer) is a tool that iteratively and randomly generates inputs to test the quality of a target program\cite{Klees2018EvaluatingTesting}. Compared to formal analysis, fuzz testing has proved to be successful in discovering critical security bugs in real software\cite{Klees2018EvaluatingTesting}. For example, \cite{wang2022automated} implemented a \ac{RRC} fuzz testing experiment for air interface protocols. Significant effort has been devoted to devising new fuzzing techniques, strategies, and algorithms. Fuzz testing has been used intensively for large-scale system cybersecurity purposes, and multi-strategies are proposed to detect cyber vulnerabilities efficiently. He et al.~\cite{he2022intelligent} proposed a state transition fuzzing framework that can be applied to different types of message identifiers. To eliminate the randomness and blindness of fuzzing, \cite{moukahal2021vulnerability} introduced a vulnerability-oriented fuzz (VulFuzz) testing framework to prioritize the fuzzing cases by security vulnerability metrics.

Even though fuzz testing can detect more emergent and unexpected behaviors than formal verification with less additional manual intervention, formal verification is much more efficient than fuzz testing with manually generalized representative mathematical expression. Leveraging the advantages of formal verification and fuzz testing has become a popular topic. In \cite{han2012mutation}, extreme cases like buffer overflow or incorrect format are discussed, combined with the advantages of protocol and mutation. Besides the extreme case, rule-based fuzzing~\cite{Salazar20215Greplay:Injection} focuses on covering all protocol-based cases. Under the limited directions defined by formal verification, coverage-guided fuzz~\cite{sheikhi2022coverage} was proposed to test the security of cyber-physical systems. Furthermore, the state-of-art vulnerability detection approach~\cite{ammann2023dy} proposed a possible combination of formal verification and fuzz testing. When extended to the long-time multi-time attacks, Ma et al.~\cite{ma2017semi} proposed a state transaction method to analyze serial attacks. Based on the formal verification, fuzz testing can efficiently locate the high-risk area. However, there are still significant gaps in highly relying on pre-assumptions of prior knowledge awareness and focusing on the specific implementation of the targeted protocols. Therefore, LZfuzz~\cite{bratus2008lzfuzz} was proposed to eliminate the requirements for access to well-documented protocols and implementations while focusing on plain-text fuzzing. And Osborne~\cite{osborne2021leveraging} et al. proposed a framework that can apply fuzz testing with area limitations in real-world experiments to narrow the fuzzing area. To address the challenges of computation power, without pre-assuming to, but leveraging on the available prior domain knowledge, we presented a multiple dimension multi-layer protocol-independent fuzzing framework in \cite{yang2023systematic}, aiming for protocol vulnerabilities detection and unintended emergent behaviors in fast-evolving 5G and NextG specifications and large-scale open programmable 5G stacks. However, these manual formal guided fuzz testing can not be automatically applied to detect cyber security vulnerabilities. In this paper, our proposed framework generates a positive feedback loop between formal verification and fuzz testing.

\subsection{Motivation}

Although fuzzing is a fast technique to detect vulnerabilities and flaws, it comes with the limitations of poor coverage which involves missing many errors and thereby limits the performance of the security vulnerabilities testing. Whereas, in formal verification methods, despite the use of abstract mathematical representations of a system under test to verify or detect the specified flaws or vulnerabilities in 5G systems, these methods inherently suffer from scalability limitations. These limitations restrict their ability to perform in more complicated and ever-larger systems due to the exponential growth of the state space with the size of the system. This limitation puts large complex systems out of the reach of the model checkers. 

Aiming for security, usability, and reliability, the objective of this system is to improve security assurance and resilience at both specification and implementations levels by discovering and mitigating vulnerabilities and unintended emergent behaviors with sufficient automation, scalability, and usability. The presented approach could be applied to various fifth-generation (5G) open programmable platforms\cite{O-RANAlliance2018O-RAN:RAN,SoftwareRadioSystems2021SrsRANSRS,Wang2021DevelopmentResearch} or other cognitive\/software defined communication systems\cite{wang2012dynamic}. 

\subsection{Contributions}
In this paper, we propose a novel approach that connects the strengths and coverage of formal and fuzzing methods to efficiently detect vulnerabilities across protocol logic and implementation stacks in a hierarchical manner.  The main contributions are listed below.

\begin{itemize}[noitemsep,topsep=5pt]
    \item We design and implement formal verification for 5G authentication and authorization specifications to detect attack traces and form attack models, which are used to guide subsequent fuzz testing and incorporate feedback from fuzz testing to broaden the scope of formal verification. 
    
    \item We perform fundamental research towards the united and amplification of formal method and fuzz testing targeting large-scale system assurance, which could benefit the interdisciplinary research community in Program Languages and Infrastructure Cybersecurity. 
    
    \item We present a proof of concept system that increases the efficiency and enables the auto-discovery of vulnerabilities and unintended emergent behaviors in 5G specifications to software stacks, and illustrated the applicability and extendability to a variety of specifications and implementations via a seven steps United Formal\&Fuzz Systematic Framework (UFSF). 
    
    
    \item Our novel approach of protocol abstraction converts the natural language-based specifications into non-ambiguous symbolic expressions, from which formal analysis models could be auto-derived. It releases the formal analysis from the labor-expertise-intensive process and leads toward the auto-formal verification. A proof of concept is performed on the \ac{NSA} 5G authentication process by converting informal protocols into a dependency table, enabling formal analysis that detects attack traces, thereby discovering $4$ attack models.
   
    
    \item By leveraging UFSF from formal verification, our integrated solution of formal guided fuzz testing further employs command-level and bit-level strategies to detect exploitable vulnerabilities and unintended emergent behaviors effectively. We successfully establish a connection among specification, implemented systems, and real-life attack models, perform a thorough examination of the complicated 5G NSA authentication and authorization process, and exploit $53$ vulnerabilities  demonstrated on srsRAN platforms. 

    \item Unlike the state-of-the-art by-piece vulnerability detection, the presented systematic vulnerability detection addressed the foundations for achieving assurance for Future G authentication and authorization in providing the panoramic vision and examination of the to-date 5G specifications. 
    
\end{itemize}

\section{System Overview}\label{system_desin}

\begin{figure}[!h]
    \centering
    \includegraphics[width=0.48\textwidth]{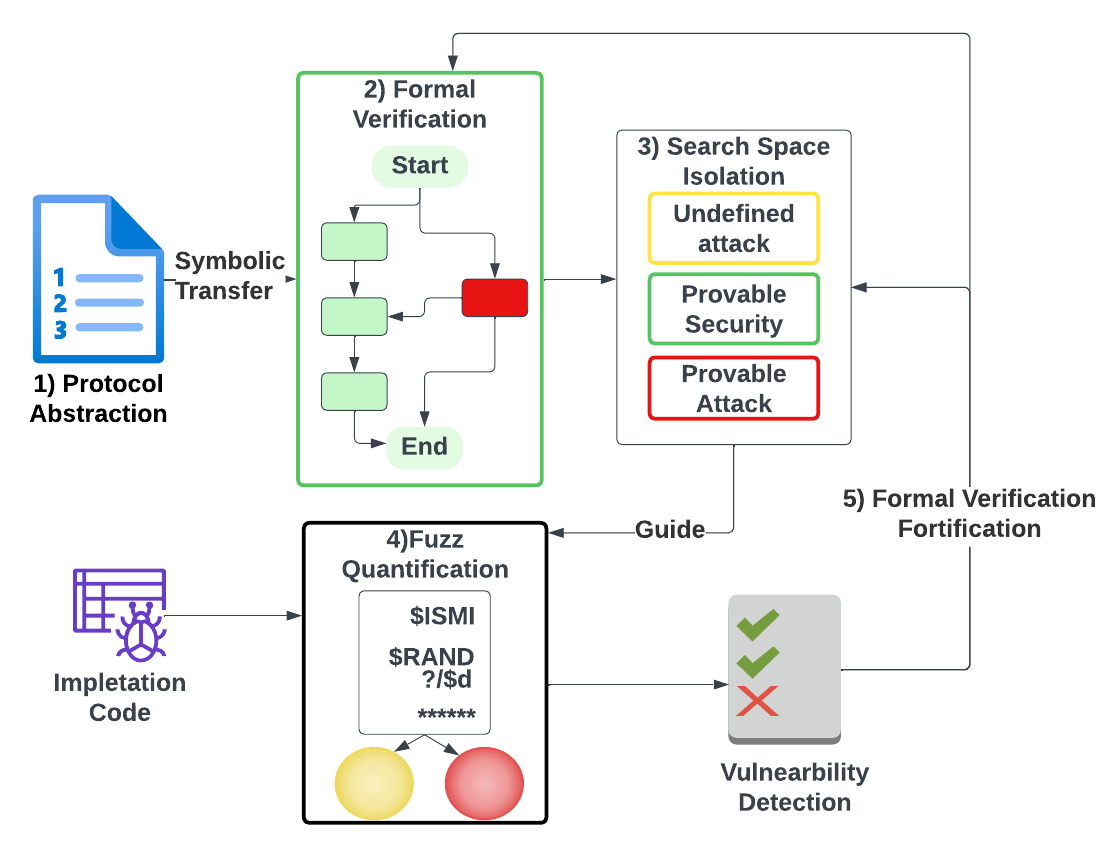}
    \caption{System Components and Connector View}
    \label{fig:formal_fuzz}
\end{figure}

Aiming at providing auto-assurance for 5G and beyond specifications to stack implementations, we present a s vulnerability and unintended emergent behaviors detection system. As shown in Fig.~\ref{fig:formal_fuzz}, The presetned system leverages the amplification and cross-validation of fuzz testing and formal verification. Our proposed framework builds up a virtuous recursive loop of the following steps:
  \begin{enumerate}[noitemsep,topsep=5pt]
        \item \textbf{Protocol Abstraction:} 
        Starting from the 3GPP technical specifications (TS) and requirements (TR), we first convert the natural language-based specifications into unambiguous symbolic expressions known as an authentication and authorization flow-graph (AAF). We then further transform this flow-graph into a properties table and generate a dependency graph. The dependency graph serves as a foundation for deriving formal analysis models automatically. This approach liberates the formal analysis process from labor-intensive and expertise-dependent tasks, facilitating auto-formal verification. It also enables incremental evolving verification by incorporating new 3GPP protocol releases into existing formal methods, eliminating the need to start the protocol abstraction process from scratch with each release.
                
        \item \textbf{Formal-based Vulnerability Detection and Attack Models:}  With the dependancy graph, we apply formal method via the ProVerif platform to conduct a logical proof of security properties and potential vulnerabilities, facilitating a robust and comprehensive evaluation of the system's security integrity. The formal method applied in the abstract protocols  not only detect the vulnerabilities in protocol design, also provide a space isolation to guide fuzz testing.  
        
        \item \textbf{Search Space Isolation:} The output of formal verification divides the search space into three sets: no vulnerabilities, attack trace detected, and uncertain areas that need further investigation. The division of the search space effectively narrows down the uncertain regions and enables the guidance and direction of fuzz testing.
        \item \textbf{Formal Guided Fuzz Testing:} With the detected attack models from formal analysis, we direct and generate a list of fuzz testing. Compared to formal analysis on the specifications, the initiated fuzz testing is performed on runtime binary systems, focusing particularly on the predefined uncertain areas and areas of identified attack traces. The guided fuzz testing is deployed to identify runtime vulnerabilities, thereby complementing the detection of vulnerabilities through logical proofs on protocols and assessing the impact of the formal detected attack models and traces. Further, it also functions as a stochastic approach for those uncertain areas that cannot be verified through formal methods.
        \item \textbf{Fortification of Protocol and Formal Verification :} Based on the vulnerabilities and unintended emergent behaviors detected by formal methods and guided fuzz testing, we derive the solutions and fortifications to either directly enhance the protocol's robustness and resilience or narrow down search spaces. By defining the space more precisely, formal verification can be further optimized, consequently extending the scope of the security assurance area.
    \end{enumerate}
    
\begin{figure}[!h]
    \centering
    \includegraphics[width=0.5\textwidth]{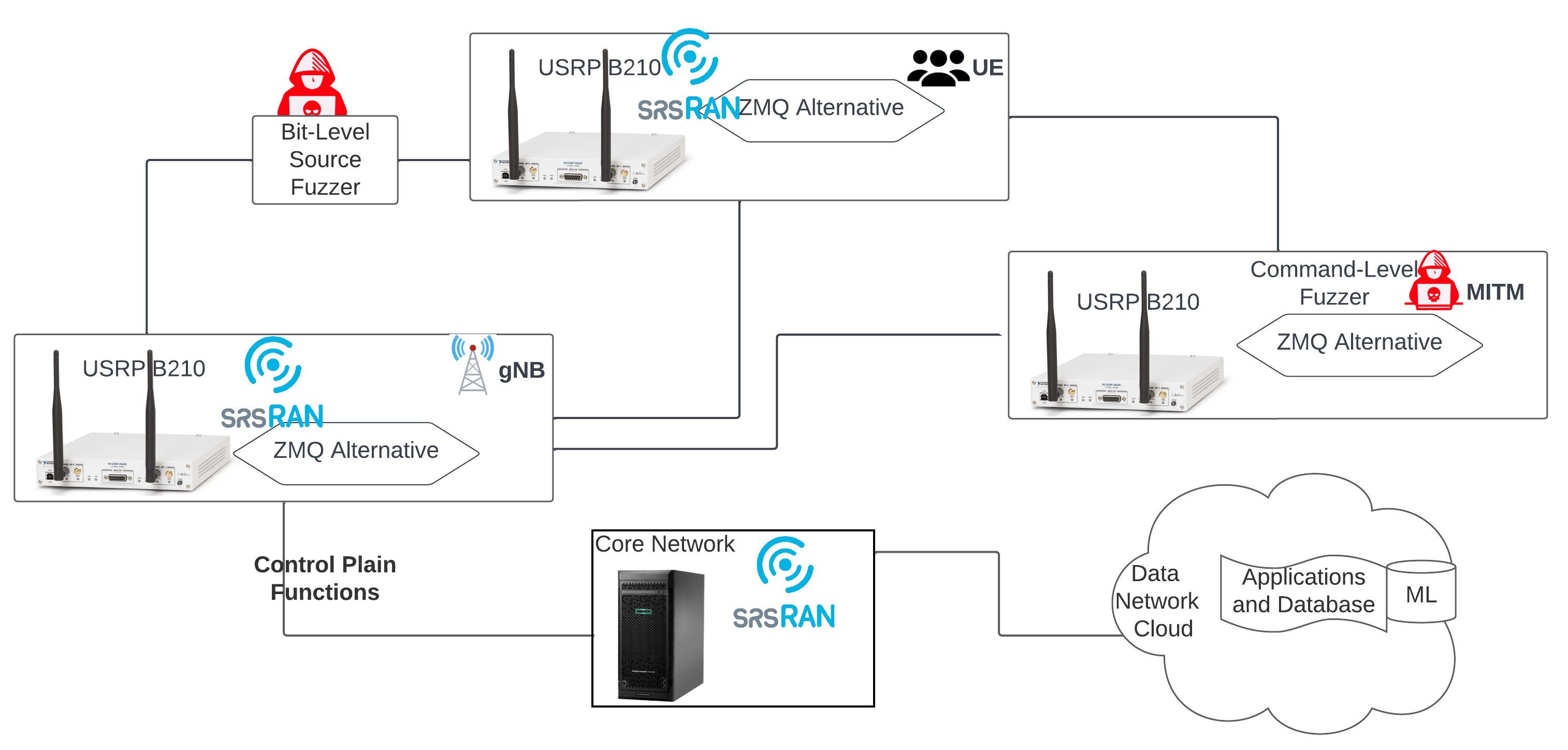}
    \caption{Experimental Platform Structure and Setup\cite{yang2023systematic}}
    \label{fig:fuzz_exp}
\end{figure}

  We further demonstrate the proposed framework by leveraging our existing platform of the fuzz testing-based digital twin \cite{JingdaYang20235GListen-and-Learn, Dauphinais2023AutomatedSystems, yang2023systematic} for 5G cybersecurity, as illustrated in Fig. \ref{fig:fuzz_exp}. Both \ac{OTA} and zeroMQ modes in legitimate communications are performed leveraging srsRAN. Interfacing with our digital twin platform, we enable mutation-based identifiers fuzzing (Bit-Level Fuzz Testing) and permutation-based command fuzzing (Command-Level Fuzz Testing) that could be used for implementation-level verification, formal discovery extension, and searching space triggering  guided from the formal method results. . With the utilization of formal result analysis,  formal guided fuzz testing, and the fortification,  our proposed framework  constructs a reinforcing loop to enhance the system's resilience

\section{Protocol Abstraction}\label{protocol_abstraction}

\subsection{Protocol and Symbolic Conversion for Formal Analysis}\label{formal_analysis}
\ac{NSA} 5G architecture can be divided into the legacy LTE authentication process and LTE-to-5G connection reconfiguration. Compared to \ac{SA} 5G network architecture, \ac{NSA} 5G architecture is more widely adopted but more vulnerable because cross-generation of protocols introduces the vulnerabilities from LTE. Therefore, we focus on the pre-authentication process of LTE in \ac{NSA} 5G architecture. As shown in Fig.~\ref{fig:lte_arch}, the LTE authentication in \ac{NSA} architecture can be divided into the following four parts:

\begin{figure*}[!t]
    \centering
    \includegraphics[width=0.9\textwidth]{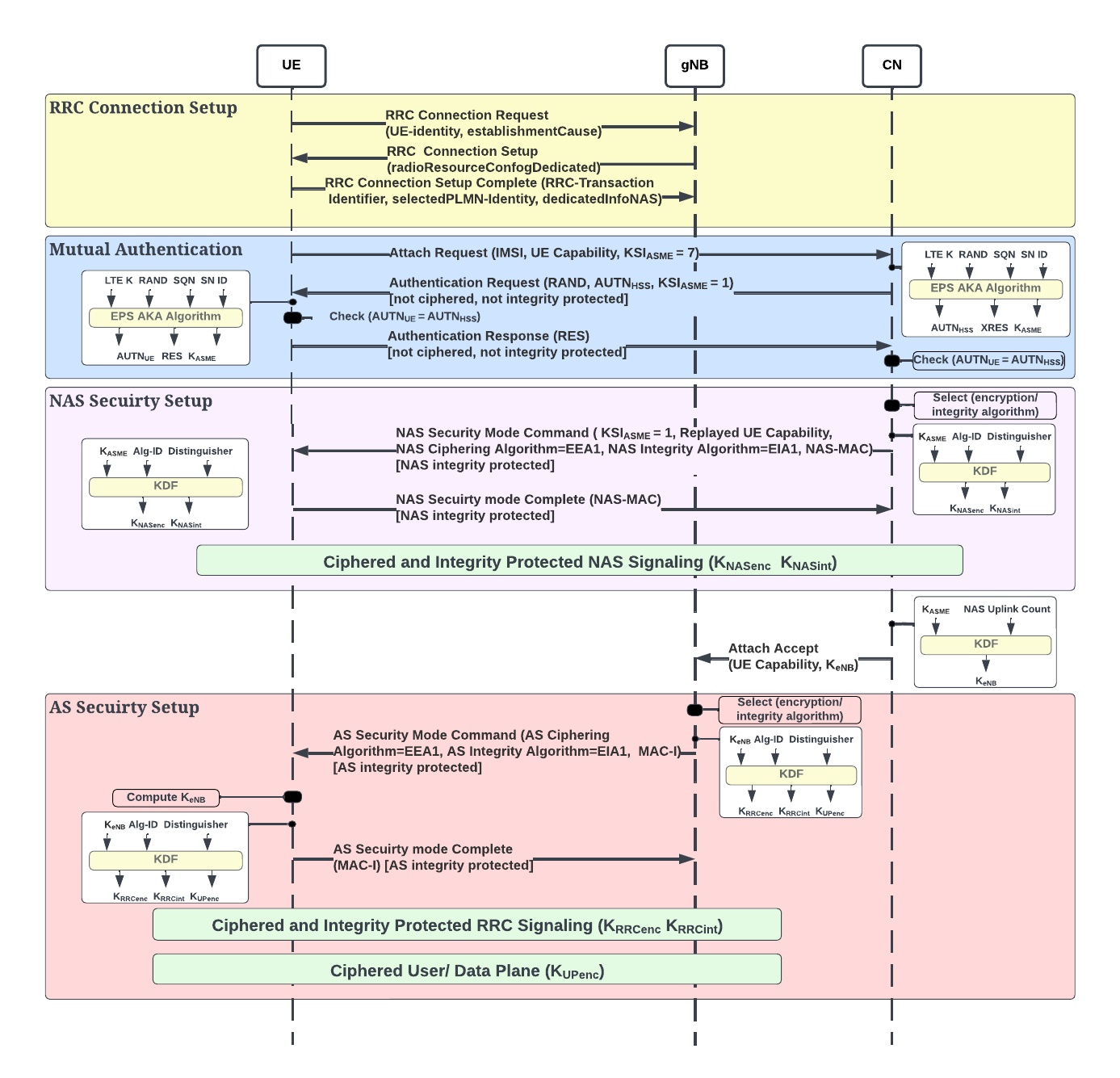}
    \caption{NSA and AS Authentication and Authorization Flowgraph.}
    \label{fig:lte_arch}
\end{figure*}

\begin{enumerate}[noitemsep,topsep=5pt]
    \item \textbf{\ac{RRC} Connection Setup:} \ac{RRC} connection setup process aims to build up connections in \ac{RRC} layer. First, \ac{UE} sends the \ac{RRC} Connection Request command with UE-identity and establishment cause to \ac{gNB}. Then, \ac{gNB} replays with radio resource configuration to \ac{UE}. If the setup process is valid, \ac{UE} will send \ac{RRC} Connection Setup Complete command with necessary identifiers to \ac{gNB} and prepare for the following \ac{NAS} security setup. In the \ac{RRC} connection setup process, we verify the reliability, consistency, and stability of communications between \ac{UE} and \ac{gNB}. Confidentiality will not be considered because the \ac{RRC} connection setup process is designed for a non-encrypted environment.
    \item \textbf{Mutual Authentication:} 
    \ac{UE} and \ac{CN} adapt \ac{EPS} \ac{AKA} algorithm as encryption and decryption tools to set up mutual authentication. In our designed formal \ac{EPS} algorithm, there are four required identifiers to get the corresponding values, $AUTN$, $RES$, and $K_{ASME}$. Even if we assume the \ac{EPS} algorithm is impregnable, the previous messages containing \ac{IMSI} and temporarily generated $rand\_id$ are neither ciphered nor integrity protected. The unencrypted mutual authentication process is vulnerable to disclosing the user identity under \ac{MITM} attacks. Based on exploited vulnerabilities and properties, we test the security impact of user identity by formal verification and simulate the \ac{MITM} attack mode.
    \item \textbf{\ac{NAS} Security Setup:} After mutual authentication, \ac{CN} needs to decide encryption algorithm and integrity algorithm. To ensure the security of \ac{NAS} communication setup, \ac{UE} and \ac{CN} communicate with integrity protection to decide encryption and integrity algorithm, and $K_{ASME}$, which is the top-level key to be used in the access network. Then \ac{UE} and \ac{CN} can get the corresponding session key for encryption and integrity of following symmetric \ac{NAS} communication.
    \item \textbf{\ac{AS} Security Setup:} \ac{NAS} security setup shares $K_{ASME}$ between \ac{CN} and \ac{UE}. However, there is still necessary to establish another channel for user status management, like \ac{RRC}. Therefore, \ac{CN} generates a key $K_{eNB}$ for \ac{gNB} based on $K_{ASME}$ and NAS up-link count and forward the $K_{eNB}$ to \ac{eNB} through the private network. Same with \ac{NAS} security setup, \ac{eNB} and \ac{UE} share the $K_{eNB}$ and selected encryption and integrity algorithm with integrity-protected communications. Then \ac{eNB} and \ac{UE} use the generated \ac{RRC} encryption key, $K_{RRCenc}$, integrity key, $K_{RRCint}$, and generated \ac{UP} encryption key, $K_{UPenc}$, to establish symmetric ciphered and integrity protected \ac{RRC} and \ac{UP} communication.
\end{enumerate}

\subsection{Properties Definition and Extraction}
Following the Flowgraph shown in Fig.\ref{fig:lte_arch}, we further extract four major security properties, confidentiality, integrity, authentication, and accounting, from the 3GPP specifications that are critical in the Formal Analysis. These four properties represent four aspects of security enhancement in the specifications: 
    \begin{enumerate}
        \item \textbf{Confidentiality} represents the ability to prevent private information from leakage.
        \item \textbf{Integrity} denotes the capability to keep the information unmodified.
        \item \textbf{Authentication} means whether the receiver can identify who and when to send the message.
        \item \textbf{Accounting} is identifying whether the current message follows the right order in session.
    \end{enumerate}

According to the four security properties, we generate an identifier-based Properties Table (PT) in Fig.~\ref{fig:dependency_table} to show the reflect the specification in the control messages. The value in the security property columns represents which identifier the current identifier depends on (N means None identifier). From Fig.~\ref{fig:dependency_table} Note 1, we conclude that \ac{RRC} connection setup process which includes three steps are unprotected in regarding confidentiality, Integrity, Authentication and Accounting. For identifiers that are protected in some properties,  we check the critical keywords\/identifiers in each propriety. The properties of the critical keywords\/identifiers become the assumptions of that property examination. For instance as shown in Note 2 of Fig.\ref{fig:dependency_table}, the integrity of $AUTN_{HSS}$ with the assumption of safe rand number (RAND) or the assumption of leaked RAND. 

The content of the Properties Table serves as the input assumption and properties in the following formal analysis. From the table, it can be observed that there exists the dependency between rows. That dependency determines the flow-graph for each formal model for vulnerability detection.    

    \begin{table*}
        \centering
        \caption{Properties Table of Protocol}
        \includegraphics[width=\textwidth]{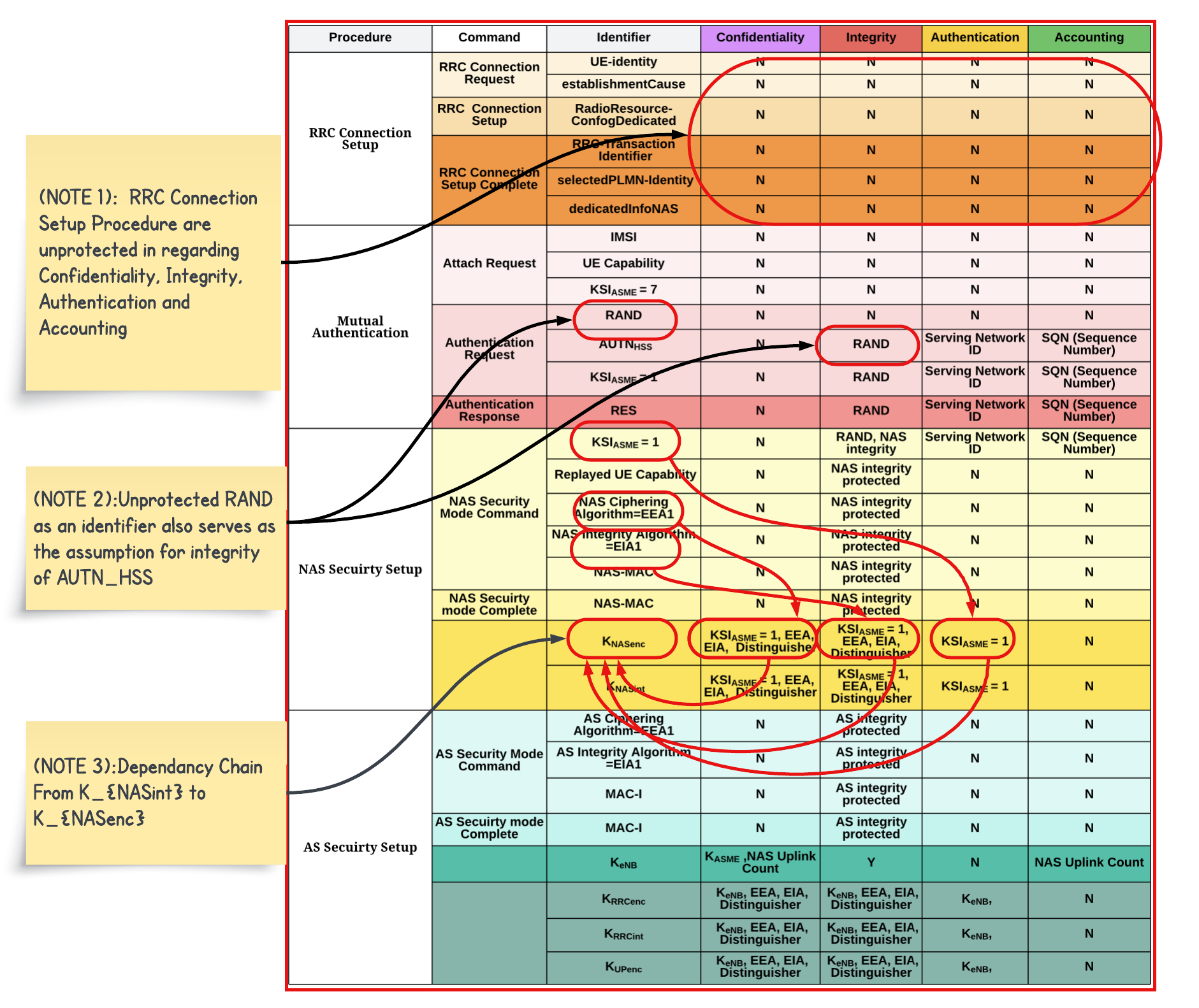}
        \label{fig:dependency_table}
    \end{table*}

    \begin{figure*}[!htb]
        \centering
        \includegraphics[width=\textwidth]{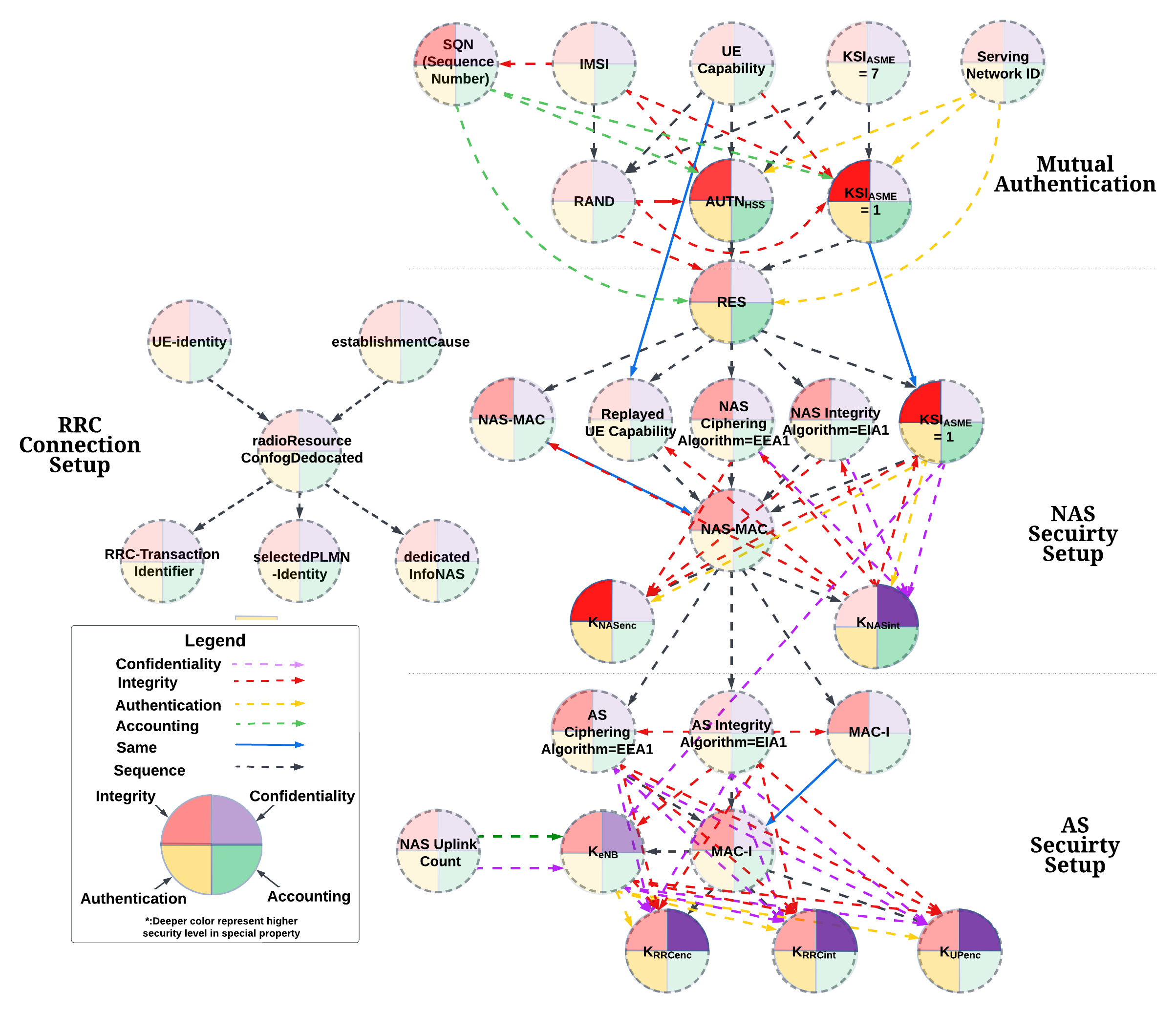}
        \caption{Dependency Graph of Protocol}
        \label{fig:dependency_tree}
    \end{figure*}

\subsection{Dependency Graph Generation}
To further visualize the dependency existed in the Properties Table, we generate a Dependency Graph(DG) in Fig.~\ref{fig:dependency_tree}. From the Dependency Graph, we can extract the dependency trace of identifiers to evaluate the chain effects along the dependency relationships and assess multi-level security risk. For example as also shown in Note3 of Fig.\ref{fig:dependency_table}, $K_{NASenc}$ have a higher integrity security level than $NAS$-$MAC$, because the integrity of $K_{NASenc}$ is protected by $NAS\;Ciphering\;Algorithm$, $NAS\;\\Integrity\;Algorithm$ and $KSI_{AMSE}$, which are protected by $K_{NASint}$, but the integrity of $NAS-MAC$ is only protected by $K_{NASint}$. Based on the security risk level, we first prove the vulnerabilities of the low-risk identifiers and then prove the security of high-risk identifiers based on the proven assumptions of low-risk identifiers. Following the security, propriety tracks provide the guidance of the test target, which narrows the target range and improves the efficiency of formal and fuzz testing.

Our security level evaluation system follows the Depth-first search (DFS) principle, and inherent the security level from the parent node (dependency node). For example, shown in~Alg.\ref{alg:security_evaluation}, the recursive algorithm adds the security level of dependent property to their security level vector. Based on different consideration and application scenarios, we use the Hadamard product of weight vector and security level vector in Equation~\ref{equ:security_level} to represent the global security level. For example, we can set weight vector as $[1,1,0.5,0.5]$ if we care more about confidentiality and integrity.
\begin{algorithm}
    \KwData{$r$ = Boolean vector of dependency relation.}
    \begin{flushleft}
        \begin{algorithmic}[1]
        \caption{Security Level Evaluation}
         \label{alg:security_evaluation}
            \PROCEDURE{Security\_Evaluation}{(node\_$v$)}
                \State [c, i, au, ac] = $[1, 1, 1, 1]$
                \WHILE{no dependent node $v^\prime$ exists}
                    \State  [c, i, au, ac] += \Call{Security\_Evaluation}($v^\prime$) $\odot$ r
                \ENDWHILE
                    
                \State \textbf{return} [c, i, au, ac]
            \ENDPROCEDURE
    \end{algorithmic}   
    \end{flushleft}
\end{algorithm}

\begin{equation}
\label{equ:security_level}
S = [\alpha_c, \alpha_i, \alpha_{au}, \alpha_{ac}] \cdot [c, i, au, ac]^T
\end{equation}

\subsection{Dependency Analysis}\label{dependency_analysis}
Based on the defined dependency graph above, we use some samples to illustrate the process mechanism of how to extract the highest risk path to the special identifier.
\subsubsection{\ac{RRC} Connection Setup Dependency Analysis}
From Fig.~\ref{fig:dependency_tree}, we conclude get that all identifiers in \ac{RRC} Connection Setup are not protected by encryption or integrity check. We can conclude that the security level of identifiers in \ac{RRC} Connection Setup$=\left[0,0,0,0\right]$.

\subsubsection{ $K_{NASenc}$ Dependency Analysis}
\begin{figure}[!t]
        \centering
        \includegraphics[width=0.5\textwidth]{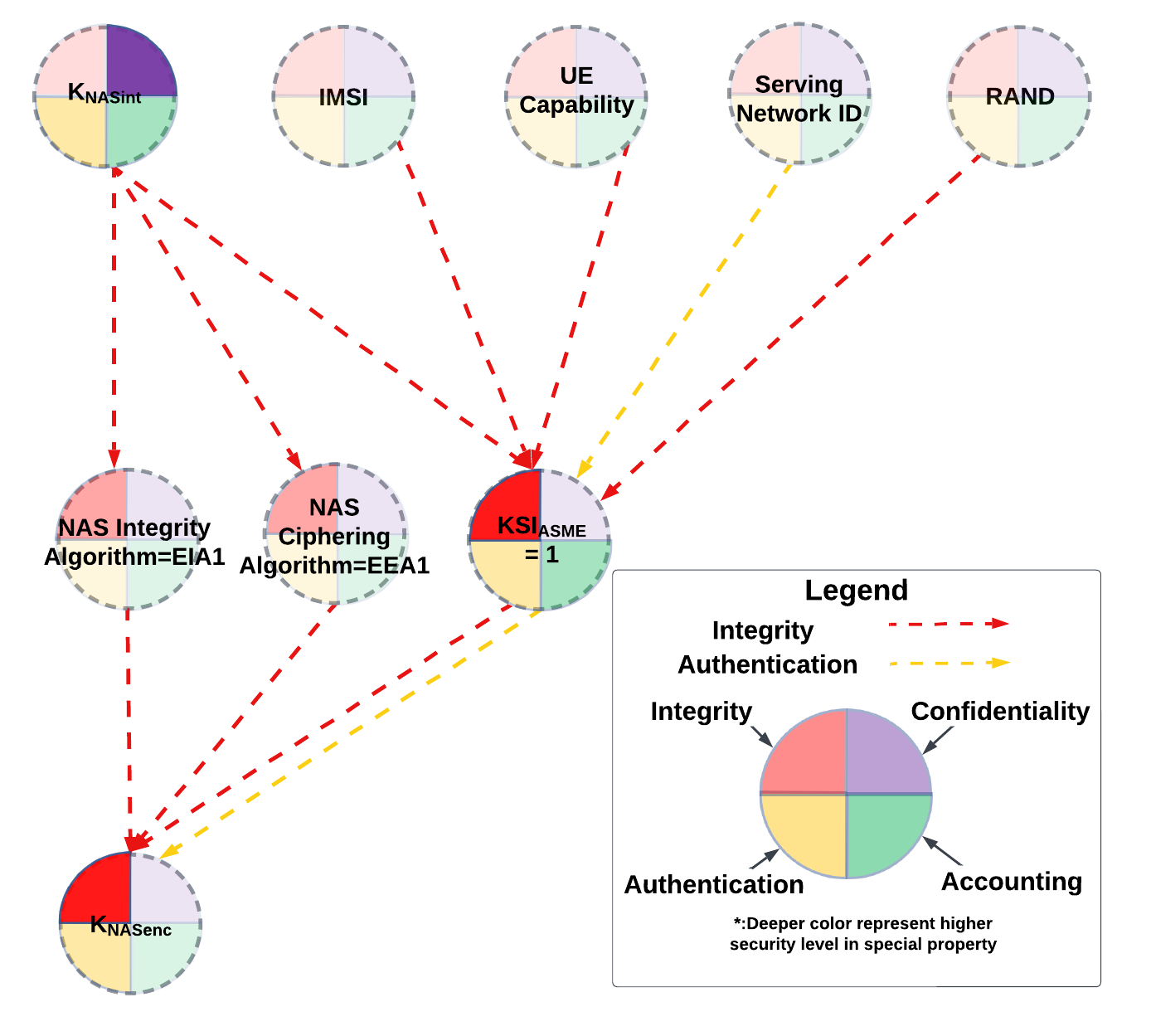}
        \caption{Dependency Relationship of $K_{NASenc}$}
        \label{fig:analysis_knasenc}
    \end{figure}

$K_{NASenc}$ is the most critical identifier in \ac{NAS} authentication process and responsible for the following \ac{NAS} communication encryption. To prove the security of $K_{NASenc}$, we extract a logical dependency graph of $K_{NASenc}$, Fig.~\ref{fig:analysis_knasenc}, from the whole dependency graph of authentication graph, Fig.~\ref{fig:dependency_tree}. From Fig.~\ref{fig:analysis_knasenc}, we can conclude that there are three direct integrity-dependent identifiers and only one direct authentication-dependent identifier. We discuss the security level from two aspects of security properties:
\begin{enumerate}[noitemsep,topsep=5pt]
    \item Authentication:
    Based on the $K_{ASME}$ derivation function, attackers can derive the $SN_id$ from the $K_{ASME}$. However, attackers can not generate the $K_{ASME}$ from the $K_{NASenc}$. Based on the authentication conduction of these three identifiers, the invertibility of the path is critical for authentication tracking. The coexistence of the authentication dependency relationship and inevitability can prove the feasibility of invertible conduction from bottom to up. 
    \item Integrity:
    The trustworthiness, consistency, and accuracy of the data throughout its life cycle is termed as integrity. Based on the dependency relationship of $K_{NASenc}$, as shown in Fig.~\ref{fig:analysis_knasenc}, only with the ability to modify three direct identifiers, secret attackers can modify $K_{NASenc}$ secretly. Furthermore, attackers can modify three direct identifiers only when they can modify all five second-level identifiers, which are directly connected to three direct identifiers. We can conclude that the minimum requirement of $K_{NASenc}$ modification is $5$ identifiers in $3$ command, including Attach Request, Authentication Request, and NAS Security Mode Command. 
\end{enumerate}
From the above proof, we can get the security level of $K_{NASenc}=\left[0,5,1,0\right]$.

\section{Formal-based Vulnerability Detection and Attack Models}\label{attack_model}

Based on the 5G authentication and authorization specification abstraction in Sec.\ref{protocol_abstraction}, we deploy formal models and analysis to describe the logical attack models and detect potential attack traces. In the ensuing section, we present four samples of vulnerabilities detection at disparate stages of the \ac{NSA} 5G authentication process and analyze the mechanisms of the exploited attack traces:(1)User Credentials Disclosure; (2)\ac{DoS} or Cutting of Device using Authentication Request,Exposing $K_{NASenc}$ and $K_{NASint}$; (3)Exposing $K_{RRCenc}$, (4)$K_{RRCint}$ and $K_{UPenc}$   . Our key findings are encapsulated in Table~\ref{tab:summary} in the result Section \ref{findings}.

\subsection{User Credentials Disclosure}\label{mutual_disclosure}
    In this attack, the adversary can exploit the transparency of \ac{RRC} Connection Setup process to effortlessly access critical user identity information, which includes but is not limited to the \ac{UE} identity and establishment cause. This illicit access enables the adversary to acquire user information and use the ensuing session key for nefarious activities such as eavesdropping and manipulation of subsequent communications.
    
    \textbf{Assumption.} The adversary can exploit the transparency of \ac{RRC} Connection Setup process to directly access any identifier within the message. Furthermore, the adversary is also capable of establish a fake \ac{UE} or a \ac{MITM} relay to eavesdrop and manipulate the messages within the \ac{RRC} Connection Setup process. To verify the security properties of identifiers within the \ac{RRC} Connection Setup process, including aspects such as confidentiality and consistency, we converted the aforementioned assumptions into ProVerif code.

    \textbf{Vulnerability.} As depicted in Fig.~\ref{fig:lte_arch}, the \ac{UE} initiates the process by sending an \ac{RRC} connection request to the \ac{CN}. Upon receiving this request, the \ac{CN} responds by transmitting the $radioResourceConfigDedicated$ back to the \ac{UE}. The \ac{UE}, in turn, obtains authentication from the \ac{CN} and responds with the $RRC-Transaction Identifier$, $selectedPLMN-Identity$ and $dedicatedInfoNAS$ to finalize the \ac{RRC} connection setup.
    Nevertheless, this process presents an exploitable vulnerability as an adversary can access all message identifiers. Such unprotected identifiers run the risk of being eavesdropped upon and modified, potentially enabling the adversary to orchestrate a \ac{MITM} relay attack.

\begin{figure}[!t]
        \centering
        \vspace{-10pt}
        \includegraphics[width=0.5\textwidth]{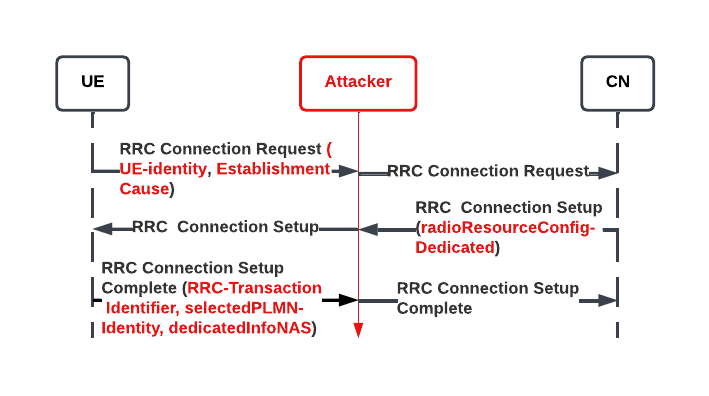}
        \vspace{-15pt}
        \caption{User Credentials Disclosure}
        \label{fig:user_identity}
        \vspace{-15pt}
    \end{figure}

    \textbf{Attack Trace Description.} Employing formal verification, we analyzed the confidentiality of identifiers within the \ac{RRC} Connection Setup process. Through this methodical investigation, we identified two categories of identifiers with the most significant impact: user identities and \ac{RRC} configuration identifiers. As illustrated in Fig.~\ref{fig:user_identity}, an attacker can access the identifiers marked in red, delineating the pathway of the attack.
    In the initial scenario, an adversary with the access to the user identity, like $UE-identity$, is capable of launch \ac{DoS} attack with real $UE-identity$. Contrary to traditional \ac{DoS} attacks, which aim to overwhelm a system's capacity, an $UE-identity$-based \ac{DoS} attack efficiently disrupts the \ac{CN} verification mechanism through repeated use of the same $UE-identity$, leading to authentication confusion.
    And in second case, with computationally derived $RRC-Transaction Identifier$, the adversary can establish a fake base station or perform a \ac{MITM} relay attack by manipulating these identifiers. In the latter case, the adversary positions between the \ac{UE} and the \ac{CN}, intercepting and modifying communications in real-time. Consequently, this attack model presents a severe threat to the security and integrity of the mobile network's communication.
        

\subsection{\ac{DoS} or Cutting of Device using Authentication Request}\label{mutual_auth}
    In the mutual authentication process, not only Attach Request command sent from \ac{UE} is neither ciphered nor integrity protected, but the Authentication Request command sent from \ac{CN} is also. Attackers can directly record and replay commands to cut off \ac{UE}. 
    
        \textbf{Assumption.}
        After \ac{CN} receives the Attach Request command sent from \ac{UE}, \ac{CN} replies Authentication Request command to confirm whether \ac{UE} is going to attach to the network and share the session key. However, because the Authentication Request command is neither ciphered nor integrity protected, \ac{UE} will be hard to verify who and when send the command.
        
\begin{figure}[!t]
        \centering
        \includegraphics[width=0.5\textwidth]{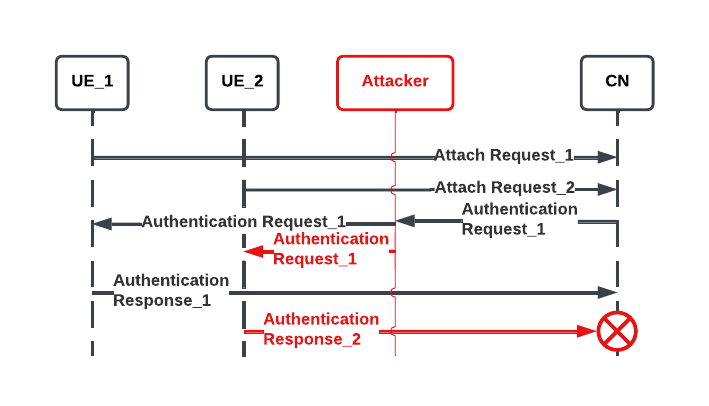}
        \caption{\ac{DoS} attack}
        \label{fig:cutting}
    \end{figure}

\begin{figure*}[!htb]
        \centering
        \includegraphics[width=\textwidth]{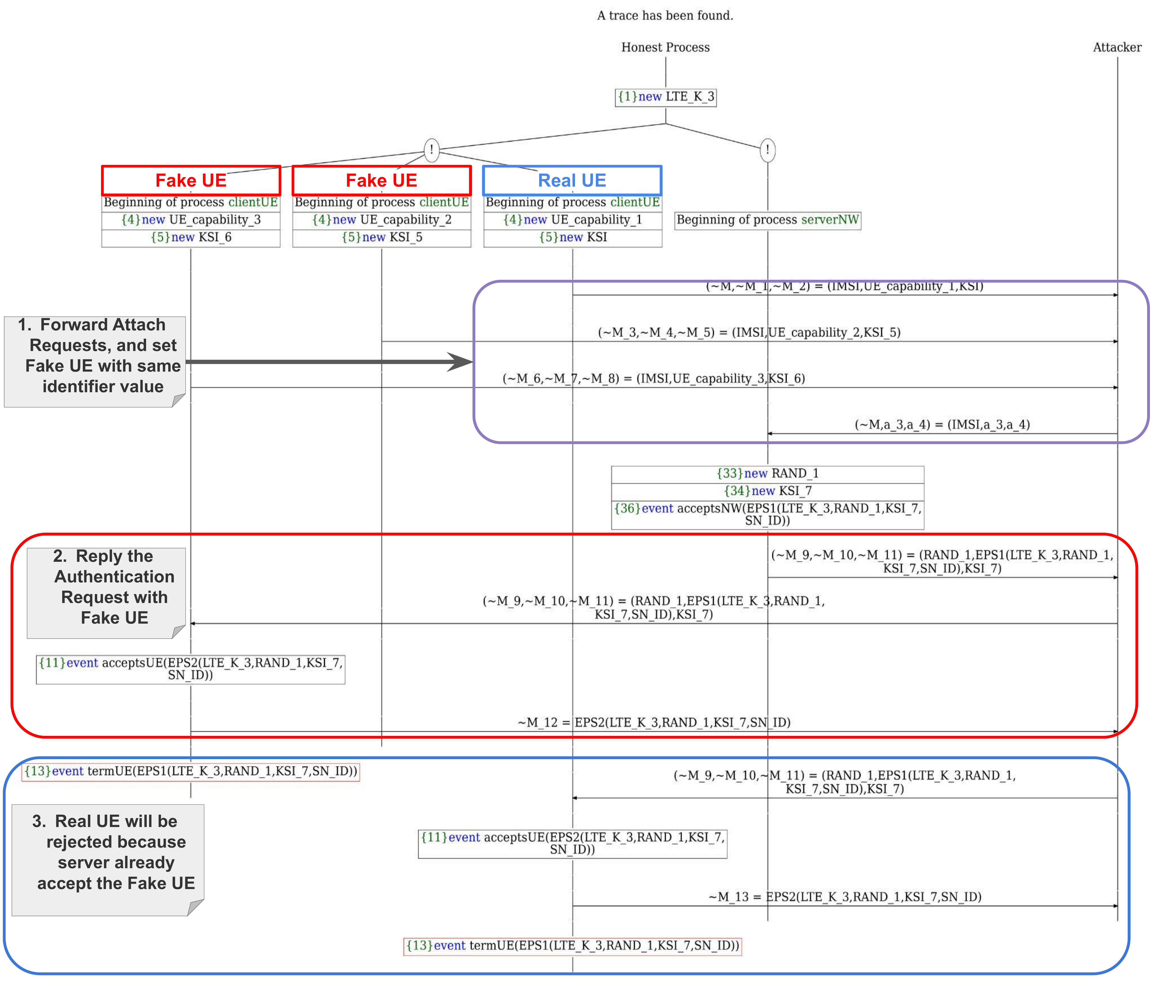}
        \caption{\ac{MITM} in Mutual Authentication}
        \label{fig:mutual_auth}
    \end{figure*}

        \textbf{Vulnerability.}
        Due to the non-confidentiality of the Authentication Request command, attackers can repeat the authentication request command to multi \ac{UE}s, as shown in Fig.~\ref{fig:cutting}. It is hard for \ac{UE} to identify which authentication request command is valid. Multi-times of authentication request command broadcasting can lead to \ac{DoS} attacks or cutting of \ac{UE}. Compared to the User Credentials Disclosure, the formal model for "\ac{DoS} or Cutting of Device using Authentication Request" is significantly more complicated. Thus, we present the formal proof of cutting off connection result shown in Fig.~\ref{fig:mutual_auth} about the interaction between 5G RAN, real-UE and fake-UE.



\subsection{Exposing $K_{NASenc}$ and $K_{NASint}$} \label{NAS}
   \ac{NAS} security establishment is only protected with integrity but not encryption, which allows attackers to access all the information but not to modify them. Attackers can fake as \ac{UE} or base station with enough information of authentication process.
    
        \textbf{Assumption.} 
        Commands of the security authentication process in \ac{NAS} security setup is only protected by $K_{NASenc}$, a key generated based on the identifiers of the first command.
        
\begin{figure}[!t]
            \centering
            \includegraphics[width=0.5\textwidth]{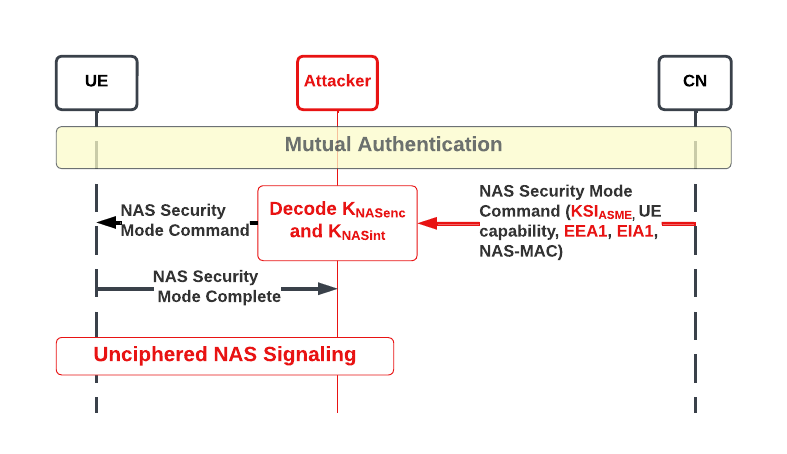}
            \caption{Exposure of \ac{NAS}}
            \label{fig:nas}
        \end{figure}

        \textbf{Vulnerability.} 
         Because commands of \ac{NAS} security mode setup are not ciphered, attackers can access the necessary identifiers and generate the corresponding session key for the following communications based on the corresponding \ac{KDF}. Then, attackers can pretend to be a base station to communicate with victim \ac{UE}, as shown in Fig.~\ref{fig:nas}. With proof of formal verification, attackers can block the communication from \ac{UE} to \ac{gNB} and continue the \ac{NAS} security setup process as the base station.
        

\subsection{Exposing $K_{RRCenc}$, $K_{RRCint}$ and $K_{UPenc}$} 
    Similar to \ac{NAS} security setup process, \ac{AS} security setup process is only integrity protected. All necessary identifiers of the following \ac{RRC} and \ac{UP} communications are transparent to attackers.
    
\begin{figure}[!h]
            \centering
            \includegraphics[width=0.5\textwidth]{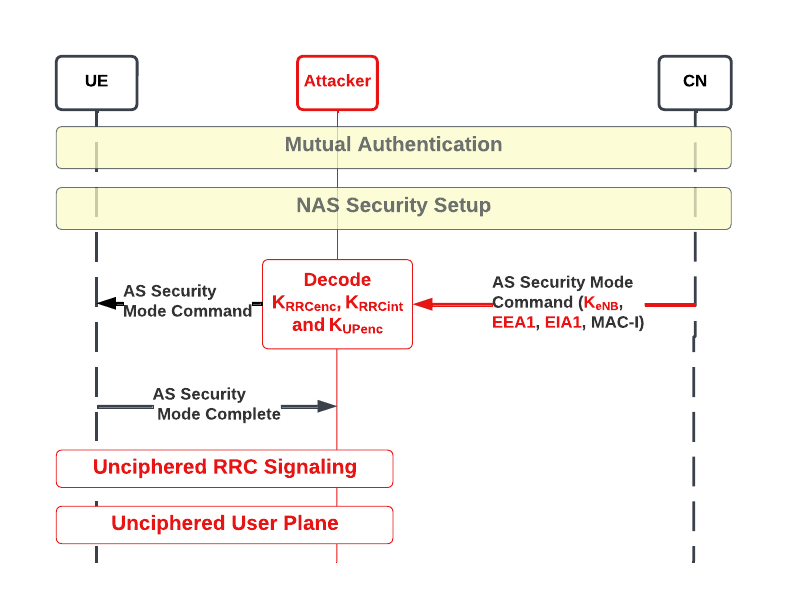}
            \caption{Exposure of \ac{AS}}
            \label{fig:as}
        \end{figure}

        \textbf{Assumption.} Similar to \ac{NAS} security setup process, all commands of \ac{AS} security setup process are only integrity protected without encryption. Attackers can generate \ac{RRC} and \ac{UP} session keys based on eavesdropped identifiers, like Fig.~\ref{fig:as}.
        
        \textbf{Vulnerability.} Based on the eavesdropped $K_{RRCenc}$, $K_{RRCint}$ and $K_{UPenc}$, attackers can monitor, hijack, and modify the commands between \ac{UE} and \ac{CN}.
        

\section{Search Space Isolation}

    \begin{figure}[!h]
        \centering
        \includegraphics[width=0.45\textwidth]{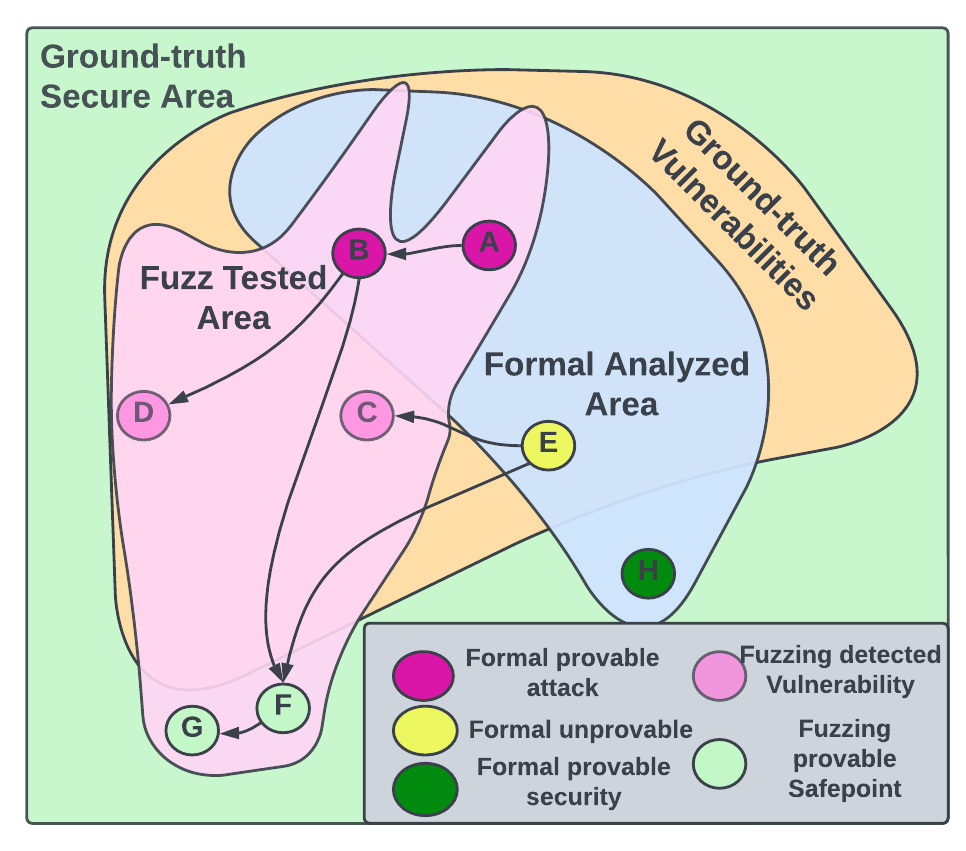}
        \caption{Definition of vulnerability region}
        \label{fig:region}
    \end{figure}

The output of formal verification divides the search space into three sets: no vulnerabilities, attack trace detected, and uncertain areas that need further investigation. The division of the search space effectively narrows down the uncertain regions and enables the scalability of vulnerability detection. Fig.~\ref{fig:region} is the visual representation of the vulnerability space. The blue area indicates the formal converted areas. Based on the conclusion from formal analysis, some traces are formally provable secure, represented by green sets in Fig.~\ref{fig:region}, and some traces are provable attacks, characterized by dark purple sets, and there is attack variance, represented by yellow sets, which are not provable by formal methods. In addition, large spaces cannot be converted by the formal method, including implementation errors and non-logical describable areas, or spaces that could be more labor-intensive and impractical to perform formal analysis. 


Thus, we introduce fuzz testing to connect with and be guided by the formal result. The formal guided fuzz testings function for two purposes: 
\begin{itemize}
    \item Compensate for areas that remain uncovered by formal verification.
    \item Evaluate the potential risks and impacts of the formal provable attack sets. 
    \item Detect identifier level unintended emergent behaviors.
\end{itemize}

\section{Formal Guided Fuzz Testing}\label{fuzz_test}
As detailed in Section~\ref{attack_model}, formal verification divided the system's security landscape into three zones: safe, non-safe, and unprovable. While the safe area necessitates no further scrutiny, the non-safe and unprovable areas warrant further investigation using fuzz testing. Specifically, we leverage fuzz testing to evaluate the risks of impact of the non-safe areas within implementation stacks, as well as to ascertain the security level within the regions previously undetermined. By leveraging our previously developed viFuzzing platform~\cite{Yang2023SystematicImplementations}\cite{JingdaYang20235GListen-and-Learn}\cite{Dauphinais2023AutomatedSystems} that enables bit-level and command-level fuzz testing for 5G and Beyond protocols and implementation stacks, we effectively perform formal guided fuzz testing and demonstrate in the range described in Fig.\ref{fig:region}. In this session, we present two sets of bit-level fuzzing and nine sets of command-level fuzzing to illustrate the operation of our formally guided fuzzing framework. 

\begin{figure}[!h]
    \centering
    \includegraphics[width=0.5\textwidth]{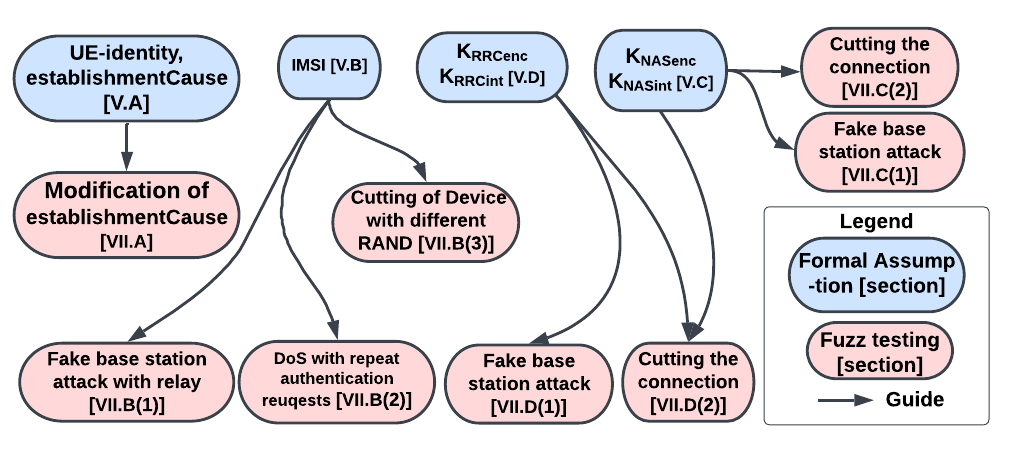}
    \caption{Integrated Solution of Formal and Fuzz Testing}
    \label{fig:formal_guide_fuzzing}
\end{figure}

We set up a relay attack mechanism interfacing our developed platform viFuzzing and srsRAN\cite{gomez2016srslte} following the attack traces detected by formal verification. The overview structure of the framework that implements formal guided fuzz testing is shown in Fig.~\ref{fig:formal_guide_fuzzing}, which illustrates the dependency and flowgraph between formal verification detections and fuzz testing results. We further present the formal guided fuzz testing cases that addressed the four  detected vulnerabilities using formal analysis in Sec.\ref{attack_model}. 

    \subsection{Modification of $EstablishmentCause$}\label{fuzz_rrc}
    Based on the proved result of formal verification, we fix the value of C-RNTI and replay the \ac{RRC} connection request commands with different values of identifier $EstablishmentCause$. Through the fuzzing result from Table.~\ref{tab:rrc_request}, modification of $EstablishmentCause$ can lead to the expected result from formal verification, but the modification of UE-Identity can not affect the connection as expected. We prove that the implementation of the srsRAN~\cite{gomez2016srslte} platform prevent some vulnerabilities of \ac{NSA} 5G communication protocol.

    Besides bit-level fuzzing, we also use command-level fuzzing to test the vulnerability of 
    incarceration with $rrc_{reject}$ and $rrc_{release}$~\cite{Hussain20195Greasoner:Protocol}. When we fixed the C-RNTI, we found the reply with $rrc_{reject}$ and $rrc_{release}$ can lead to disconnection and repeat $rrc_{reject}$ and $rrc_{release}$ can lead to failed connections.
    
        \begin{table}
        \centering
        
\caption{Fuzzing Result of $establishmentCause$ Modification}
\includegraphics[width=0.5\textwidth]{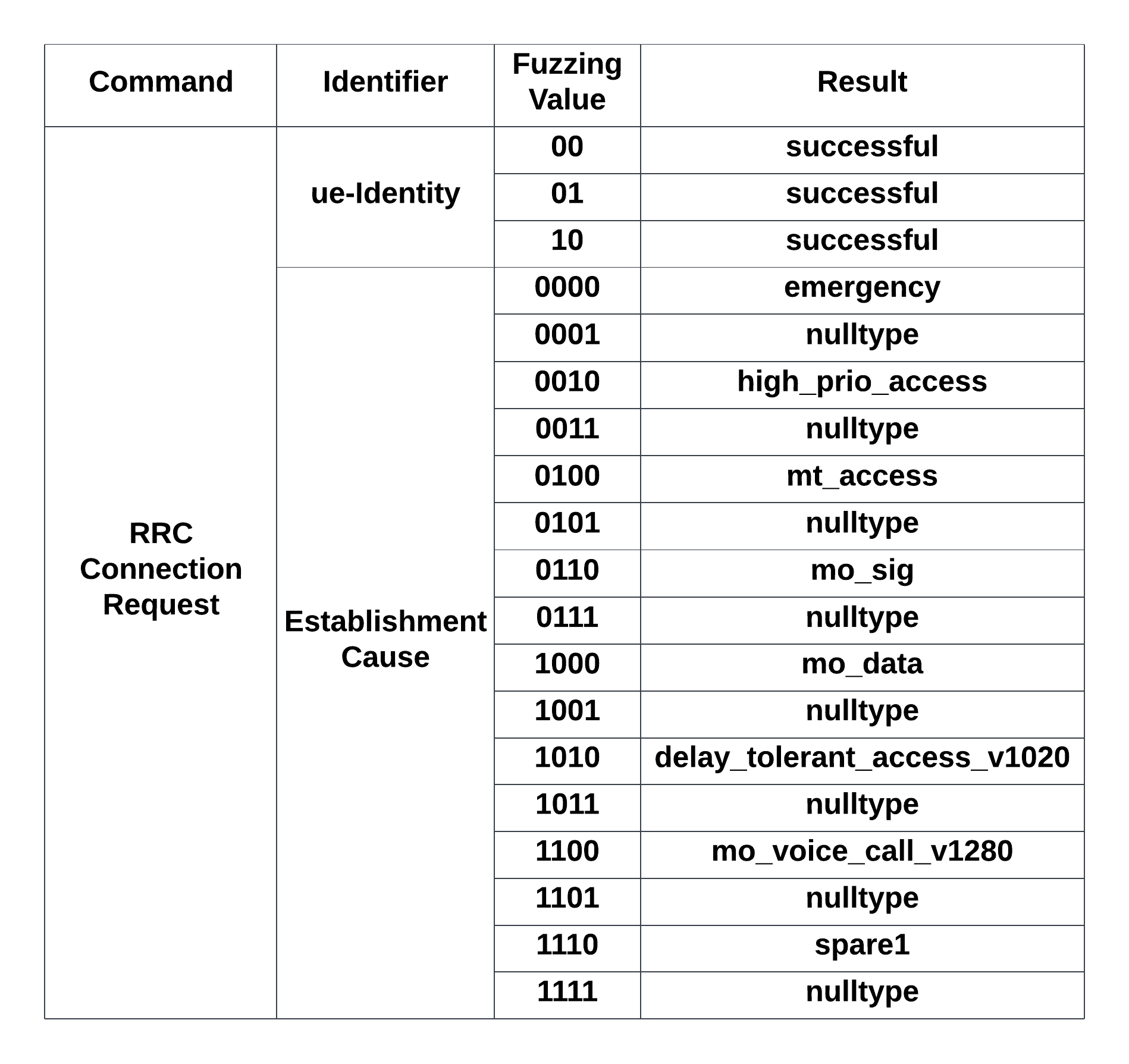}

        \label{tab:rrc_request}
        \end{table}
    \subsection{Repeat Authentication Request Command}
    Based on Section~\ref{mutual_auth}, the attacker can disconnect multi \ac{UE}s with the repeat of Authentication Request. Therefore, in our fuzzing attack model, the attacker can record the Authentication Request command from one \ac{UE} and forward the recorded Authentication Request command to other \ac{UE}s. To verify the performance of the fuzzing framework, we set up three following scenarios:
        \begin{enumerate}
            \item \textbf{Only attacker can send command to \ac{UE}.} In this case, \ac{UE} replies authentication response and try to establish a connection, which proves what we found in Section~\ref{mutual_auth} by the formal method.
            \item \textbf{One \ac{CN} and multi attackers compete to send same command to \ac{UE}.} Even if \ac{UE} gets confused by multi-times of authentication requests, \ac{UE} still has the ability to reply by sending an authentication response to \ac{CN}.
            \item \textbf{One \ac{CN} and multi attackers compete to send different command to \ac{UE}.} In this scenario, while attackers use different RAND and disclosure \ac{IMSI} to generate different Authentication Request commands and forward different commands to \ac{UE}, \ac{UE} is more likely to reply to the attackers' requests.
        \end{enumerate}

    \subsection{Exposure of $K_{NASenc}$ and $K_{NASint}$}
     From Section~\ref{NAS}, we can conclude that the attacker in the \ac{MITM} relay model has the ability to act as either \ac{UE} or \ac{CN}. Compared to complex initial steps in the traditional fuzz testing model, our proposed fuzzing framework only needs a few steps to prove the feasibility and detect the implementation vulnerabilities. We  illustrate the detailed fuzzing implementation based on formal assumptions in the following:
     \begin{enumerate}[noitemsep,topsep=5pt]
         \item \textbf{\ac{MITM} attack as fake base station.} Unlike the traditional fuzz testing approach, our framework can do fuzz testing with only access to communicated commands. The following steps illustrate the process flow of our novel proposed framework:
         \begin{itemize}[noitemsep,topsep=5pt]
             \item First, our framework records normal communication commands.
             \item Then, our framework forwards the commands between \ac{UE} and \ac{CN} as normal until mutual authentication establishment with fixed same \ac{IMSI} and RAND.
             \item After mutual authentication is established, our framework intercepts the commands from \ac{UE} and reply with corresponding commands based on the record communication history.
         \end{itemize}  
         The result proves that attackers have the ability to deploy \ac{MITM} attack as the fake base station.
         \item \textbf{Cutting the connection between \ac{UE} and \ac{CN}.} Besides fuzz testing of the fake station with blocked signals, our framework can verify the feasibility of signal competition. The detailed process is listed as follows:
         \begin{itemize}[noitemsep,topsep=5pt]
             \item First, our framework records multi-times of normal communication commands with different \ac{IMSI} and RAND.
             \item Then, our framework establishs mutual authentication with another \ac{IMSI} and RAND.
             \item Unlike the previous fuzz testing case, our framework replies with corresponding commands and forwards the commands from \ac{CN}, which simulates the \ac{DoS} attack.
         \end{itemize}
         Most \ac{DoS} attacks cut off the connection between \ac{UE} and \ac{CN}. The result proves the vulnerabilities of \ac{NAS} security setup process. We can conclude the multi \ac{NAS} security mode commands attack is an efficient attack model.
     \end{enumerate}

    \subsection{Exposure of $K_{RRCenc}$, $K_{RRCint}$ and $K_{UPenc}$}
        Similar to fuzz testing on \ac{NAS} security setup, we design two kinds of fuzzing strategies:
        \begin{enumerate}[noitemsep,topsep=5pt]
            \item \textbf{\ac{MITM} attack as fake base station.} Same with \ac{NAS} fuzzing case, attackers can successfully fake as a base station when blocking the signals from \ac{CN}.
            \item \textbf{Cutting the connection between \ac{UE} and \ac{CN}.} \ac{DoS} attacks with multi times of \ac{AS} security mode commands have a high probability of cutting off the connection between \ac{UE} and \ac{CN}.
        \end{enumerate}

\section{Fortification of Protocol and Formal Verification}

Based on the results from formal analysis and guided fuzz testing, vulnerabilities detected by fuzz testing are feedback to the formal result and search space, which lead to the fortification of protocol and formal verification. This is a crucial component in improving the resilience of 3GPP specifications.  

\subsection{User Credentials Disclosure}

The adversary can exploit the transparency of \ac{RRC} Connection Setup process to effortlessly access critical user identity information, which includes but is not limited to the \ac{UE} identity and establishment cause. This illicit access enables the adversary to acquire user information and use the ensuing session key for nefarious activities such as eavesdropping and manipulation of subsequent communications.

Given the significance and susceptibility of identifiers within the \ac{RRC} Connection Setup process, it is imperative to implement integrity protection measures for the $RRC-Transaction Identifier$. Additionally, adopting a hash value approach can assist in preventing the disclosure of \ac{UE} identity, further reinforcing security measures in this critical process.

\subsection{\ac{DoS} or Cutting of Device using Authentication Request}\label{mutual_auth_forti}
    In the mutual authentication process, not only Attach Request command sent from \ac{UE} is neither ciphered nor integrity protected, but the Authentication Request command sent from \ac{CN} is also. Attackers can directly record and replay commands to cut off \ac{UE}.

Based on the analysis of detected vulnerabilities, it is necessary to develop a verification mechanism to identify the validation of commands. The encryption or integrity protection of Authentication Requests becomes necessary for mutual authentication to guarantee the security of initial identifiers for the security establishment process. Based on the principle of minimum change of the current protocol, we propose the following two solutions:
\begin{itemize}[noitemsep,topsep=5pt]
\item \textbf{\ac{EC-AKA}~\cite{6235098}.} \ac{EC-AKA} proposed new asymmetric encryption to enhance user confidentiality before symmetric encryption is determined. However, this solution increases the cost of stations like public key broadcasting.
\item \textbf{Hash value to represent \ac{IMSI}~\cite{3gpp.29.118}.} This approach can prevent attackers from getting the users' identities. However, attackers can still modify or deploy \ac{DoS} attacks.
\item \textbf{Hash value with integrity protection~\cite{khan2018defeating}.} Khan et al. proposed a combined solution, which uses hash values to represent \ac{IMSI} and adds checksum value to protect integrity. Furthermore, the following commands in the LTE security setup process can be encrypted by original \ac{IMSI}, which is invisible to the attacker but known to \ac{UE} and \ac{CN}. Hash value with integrity protection is an optimal solution that can provide enough security for user identity at a low cost.
\end{itemize}

\subsection{Exposing $K_{NASenc}$ and $K_{NASint}$} 
   \ac{NAS} security establishment is only protected with integrity but not encryption, which allows attackers to access all the information but not to modify them. Attackers can fake as \ac{UE} or base station with enough information of authentication process.

        Same with Section \ref{mutual_auth_forti}, there are two encryption methods to protect the NAS security setup:
            \begin{enumerate}
                \item Broadcasting asymmetric public key from \ac{gNB} can be applied to encrypt the commands.
                \item \ac{NAS} security setup process can encrypt with original \ac{IMSI} as symmetric key, while the hashed \ac{IMSI} is used for \ac{RRC} connection setup.
            \end{enumerate}

\subsection{Exposing $K_{RRCenc}$, $K_{RRCint}$ and $K_{UPenc}$} 
Similar to \ac{NAS} security setup process, \ac{AS} security setup process is only integrity protected. All necessary identifiers of the following \ac{RRC} and \ac{UP} communications are transparent to attackers.

As proposed in previous sections, we can use asymmetric encryption to cipher the communicated commands between \ac{UE} and \ac{gNB}. And we also can use hashed \ac{IMSI} as the symmetric key to encrypt the commands.

\section{Result Analysis and Performance Assessment}\label{comparsion}

\subsection{Vulnerability Findings via Formal Method and Guided Fuzz Testing}\label{findings}

The detailed detected attack models and vulnerabilities have been described in details in the previous sessions. The summary of the vulnerabilities findings are listed in Table \ref{tab:summary}. At the protocol level, $4$ attack model categories, including modification of Radio Resource Control (RRC) connection, Denial of Service (DoS) or device disconnection using Authentication Request, exposure of $K_{NASenc}$ and $K_{NASint}$, and exposure of $K_{RRCenc}$, $K_{RRCint}$, and $K_{UPenc}$, are extrapolated from the attack traces inferred through formal verification. Following the proposed formal guided fuzz testing framework shown in Fig.\ref{fig:formal_fuzz}. In bit-level guided fuzzing, our system uncovers $8$ vulnerabilities. In command-level fuzzing, our framework detected $44$ vulnerabilities. Via the systematic approach, the list of vulnerabilities and proposed solutions and fortifications significantly enhance the resilience of the 3GPP specification and large-scale implementations, like srsRAN in our demonstration. More importantly, unlike the state-of-the-art by-piece vulnerability detection, it addressed the foundations for achieving assurance for Future G authentication and authorization in providing the panoramic vision and examination of the to-date 5G specifications.

\begin{table*}[]
\caption{Summary of Vulnerability Findings and Comparison with Existing Exploits}
\begin{tabular}{|m{1.7cm}|m{2.5cm}|m{1.7cm}|m{1.5cm}|m{2.5cm}|m{2.8cm}|m{1.8cm}|}
\hline
\textbf{Formal Derived Attack Models } &
\textbf{Vulnerability} &
\textbf{Assumption} &
\textbf{Related Existing Exploits}&
\textbf{Solution} &
\textbf{Guidance to fuzz} &
\textbf{Executable Vulnerabilities via Guided Fuzzing}
\\ \hline
  Modification of \ac{RRC} Connection &
  Modified commands can disable the \ac{RRC} functions &
  known C-RNTI or TMSI  &
  Related to ~\cite{Hussain20195Greasoner:Protocol} &
  Integrity protection
  &
  Fuzz testing can start with different \ac{RRC} status. &
  \begin{center}54\end{center}
  \\ \hline

 \ac{DoS} or Cutting of Device using Authentication Request. &
  \ac{UE} accepts authentication request without integrity.&
  None &
  Related to \cite{tsay2012vulnerability}\cite{6235098}\cite{3gpp.29.118}\cite{khan2018defeating} &
        \begin{itemize}[leftmargin=*]
          \item \ac{EC-AKA}~\cite{6235098} 
          \item Hashed \ac{IMSI}~\cite{3gpp.29.118}
          \item Hashed \ac{IMSI} with integrity check~\cite{khan2018defeating}
      \end{itemize}
  & 
  Repeat authentication request commands can be fuzzed at random time to test \ac{DoS} and cutting of device attack. &
  \begin{center}3\end{center}
  \\ \hline

  Exposing $K_{NASenc}$ and $K_{NASint}$ &
  All \ac{NAS} information can be monitored, hijacked and modified. &
  known \ac{IMSI}, \ac{MITM} relay &
  Related to \cite{raza2018exposing} &
  \begin{itemize}[leftmargin=*]
    \item Asymmetric encryption
    \item Hashed \ac{IMSI} based encryption
  \end{itemize}&
  \ac{NAS} fuzz testing can start with known $K_{NASenc}$ and $K_{NASint}$. &
  \begin{center}2\end{center}
  \\ \hline
  Exposing $K_{RRCenc}$, $K_{RRCint}$ and $K_{UPenc}$ &
  All \ac{RRC} and UP information can be monitored, hijacked and modified. &
  known \ac{IMSI}, MITM relay &
  New Discovery &
  \begin{itemize}[leftmargin=*]
    \item Asymmetric encryption
    \item Hashed \ac{IMSI} based encryption
  \end{itemize} &
  \ac{RRC} fuzz testing can start with known $K_{RRCenc}$ and $K_{RRCint}$; UP fuzz testing can start with known $K_{UPenc}$. &
  \begin{center}2\end{center}
  \\ \hline
\end{tabular}

\label{tab:summary}
\end{table*}




\subsection{System Assessment of Computation Complexity in Formal Guided Bit-Level Fuzzing}

Fuzz testing is a systematic brute-force vulnerability detection approach that involves providing large amounts of random data to find security vulnerabilities. However, it is not computationally feasible to complete vulnerability detection for the whole 5G \ac{NSA} protocol, even for a single command. State of the art  rule-based bit-level fuzz testing strategy has been proposed, such as~\cite{Salazar20215Greplay:Injection}, which narrows the scope of fuzz testing to specific identifiers by following the protocol rules. Although the rule-based mutation fuzz testing strategy achieves an order of magnitude reduction in computational complexity, there are still meaningless randomly generated inputs. Our proposed formal-guided fuzz testing strategy follows formal verification assumptions and generates three sets of a few representative inputs: formal-based legal inputs, formal-based illegal inputs, and randomly generated inputs. Formal-based inputs must follow the protocol-defined rules or format, but not randomly generated inputs. 


One of the novelties and advances lies in the scalability of our proposed system as the number of commands increases in complex protocols. To verify complex protocols via formal methods, formal analysis requires significant manpower and computational power. Meanwhile, attempting to cover the entire space via fuzz testing in the current state-of-the-art methodology requires an enormous number of test cases and impractical computation time, as the size of fuzz testing in the brute fuzzing strategy exhibits exponential growth. On the contrary, our presented formal-guided fuzz testing approach maintains linear growth as the number of commands increases. In this session, we perform a quantitative comparison between brute force fuzz testing, state of the art bit-level fuzzing, and the formal guide fuzz testing. 

As depicted in Eq.~\ref{equa:burte_force}, the brute-force fuzzing strategy indiscriminately flips bits within randomly selected command sets. Conversely, the rule-based fuzzing strategy~\cite{Salazar20215Greplay:Injection}, as expressed in Equation~\ref{equa:rule_based}, confines bit modifications to the identifiers within randomly chosen command sets. In contrast to these approaches, our formal-guided fuzz testing identifies the bit-level fuzzing command first. Focusing on the target commands and restricts alterations to various types of identifiers, as elucidated in Eq.~\ref{EQ-4}.

For the brute force fuzz testing complexity: 

\begin{equation}
\label{equa:burte_force}
N_{brute\_force} = 2^{\sum_{k=0}^K{\lvert c_k \lvert}}
\end{equation}
where
$N_{brute\_force}$ denotes the number of fuzz testing cases via Brute Force. $C = [c_1, c_2,  \cdot \cdot \cdot,c_K ]$ is the sets of potential commands in the target procedures fuzz testing. $K$ represents the number of target commands in fuzz testing, whereas $\lvert c_k \lvert$ is the number of bits in command $k$.

For state of the art rule-based fuzz testing complexity: 
\begin{equation}
\label{equa:rule_based}
N_{rule\_based} = 2^{\sum_{k=0}^K{\lvert c\_I_k \lvert}}
\end{equation}
where $c\_I_{k,i} \in [ c\_I_{k,1},c\_I_{k,2}, \cdot \cdot \cdot, c\_I_{k,i} ]$ represents the identifier sets in command $c_k$ and $\lvert c\_I_{k,i} \lvert$ is the number of bits in identifier $i$ in command $c_k$, $\sum_{i=1}^{I_k}\lvert c\_I_{k,i} \lvert \ll \lvert c_k \lvert$ where $I_k$ is the number of identifiers in command $c_k$.

For formal guided fuzz testing complexity: 

\begin{equation}
N_{formal\_guided} = \sum_{k=1}^{T}\sum_{j=1}^{\lvert ct\_I_k \lvert}type(ct\_I_{k,j})
\label{EQ-4}
\end{equation}
where $T = \lvert C\_target \lvert $ is the number of target commands, whereas $C\_target = [ct_1, ct_2, \cdot \cdot \cdot, ct_T]$. It is to be noted that $C\_target$ represents a subset of commands that were detected by a formal analysis as vulnerable commands that needed to be tested with fuzzing, that is, $C\_target\subseteq [c_1, c_2,  \cdot \cdot \cdot,c_K ]$. $\lvert ct\_I_{k}\lvert$ denotes the number of identifiers in target command $ct_k$. $ct\_I_{k,j}$ is the identifier $j$ of target command $ct_k$, while $type(ct\_I_{k,j})$ is the number of logical types of identifier $j$ in target command $ct_k$, including legal and valid value, legal and invalid value, and illegal random value.

\begin{figure}[!htb]
        \centering
        \includegraphics[width=0.5\textwidth]{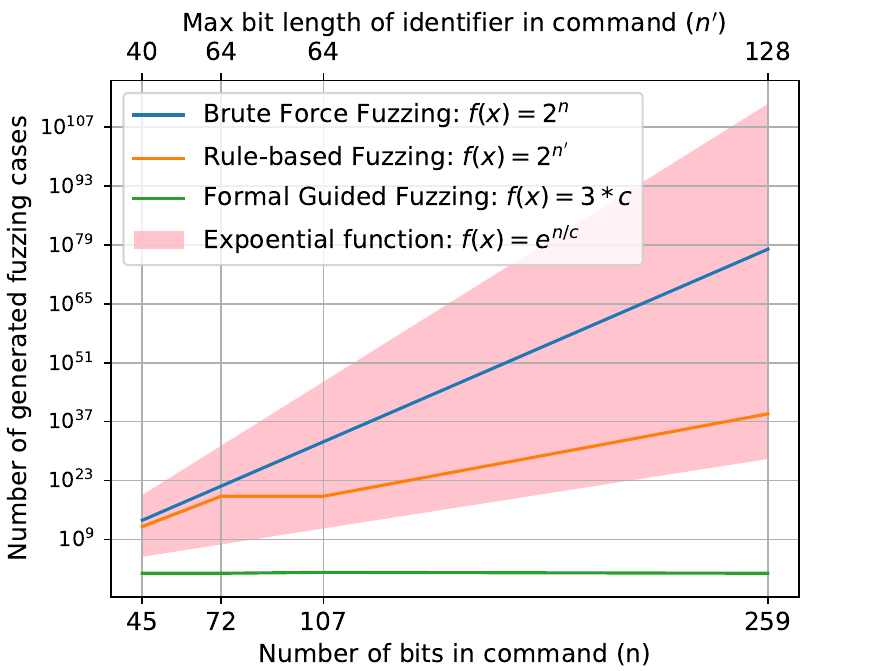}
        \caption{Comparison of Different Bit-level Fuzzing Strategy Efficiency}
        \label{fig:Comparison}
    \end{figure}

The comparison of computation complexity following Eq. \ref{EQ-4}  with $4$ fuzz strategies is shown in Fig.~\ref{fig:Comparison}, in which fuzzing strategies are selected based on various application scenarios.
\\

\textbf{(1) Connection Request command bit-level fuzzing:} Based on the guidance of formal verification in Section~\ref{mutual_disclosure}, the \ac{RRC} Connection Request command, which includes $40$ bits of UE-Identity, $4$ bits of $EstablishmentCause$, and $1$ bit of spare, is vulnerable to \ac{DoS} or \ac{MITM} attacks. Traditional brute-force fuzz testing generates more than $2^{45}$ fuzzing cases, and rule-based fuzzing generates $2^{40}+2^{4}+1$ fuzzing cases based on the defined identifiers. However, our formal guided fuzzing strategy requires only 9 fuzzing cases, including one legal UE-Identity case, one illegal UE-Identity case, one random out-of-rule UE-Identity case, $2$ legal/illegal $EstablishmentCause$ cases, $1$ random out-of-rule $EstablishmentCause$ case, one legal spare case, one illegal spare case, and one out-of-rule spare case.
\\
\noindent
\textbf{(2) Authentication Request command bit-level fuzzing:} Formal verification proved the Authentication Request command is the critical part for \ac{DoS} or fake station attacks. Inside the Authentication Request command, there are 128 bits of $RAND$, 128 bits of $AUTN_{HSS}$ and 3 bits of $KSI_{ASME}$. Our proposed formal guided fuzzing strategy generates $3\times3$ fuzzing cases, while brute-force fuzzing generates $2^{259}$ cases and rule-based fuzzing generates $2^{128}+2^{128}+2^{3}$ cases.
\\
\noindent
\textbf{(3) \ac{NAS} Security Mode command bit-level fuzzing:} To verify the formal assumptions in \ac{MITM} and cutting of the connection attacks, we make bit-level fuzzing on \ac{NAS} Security Mode command. \ac{NAS} Security Mode command has 3 bits of $KSI_{AMSE}$, 4 octets of \ac{UE} capability, 4 bits of $EEA1$, 4 bits of $EIA1$, and 8 octets of $NAS-MAC$. Brute force fuzzing needs all possible permutations and random inputs, at least $2^{107}$ cases. Rule-based fuzzing generates at least $2^3+2^{32}+2^4+2^4+2^{64}$ cases. However, our proposed formal guided fuzzing only needs $3\times5$ cases.
\\
\noindent
\textbf{(4) \ac{AS} Security Mode command bit-level fuzzing:} To verify the formal assumptions, \ac{AS} Security Mode command bit-level fuzzing is necessary. Similar to \ac{NAS} Security Mode command, \ac{AS} Security Mode command contains $4$ bits of EEA1, $4$ bits of EIA1, and $8$ octets of MAC-I. Like illustrated in \ac{NAS} Security Mode command bit-level fuzzing, formal guided fuzzing generates $3\times3$ cases. In contrast, brute force fuzzing generates at least $2^{72}$ cases, and rule-based fuzzing generates $2^4+2^4+2^{64}$ cases.


Fig.~\ref{fig:Comparison} provides an intuitive visualization that compares the effectiveness of different fuzzing strategies. The upper and lower bounds of the pink area are represented by values of "c=1" and "c=4" in Fig.~\ref{fig:Comparison}. Notably, it is evident that brute force fuzzing and rule-based fuzzing exhibit exponential growth patterns. In contrast, our proposed formal guided fuzzing approach demonstrates linear growth, requiring considerably less computational power for vulnerability verification and localization. 
The superiority of our method in terms of efficiency and scalability enables a realistic testing and vulnerability detection across the entire specifications, and provides the assurance and confidence in 5G system, especially when applied to the critical infrastructures. 


\subsection{System Assessment of Computation Complexity in Formal Guided Command-Level Fuzzing}\label{efficiency}

\begin{figure}
    \centering
    \includegraphics[width=0.5\textwidth]{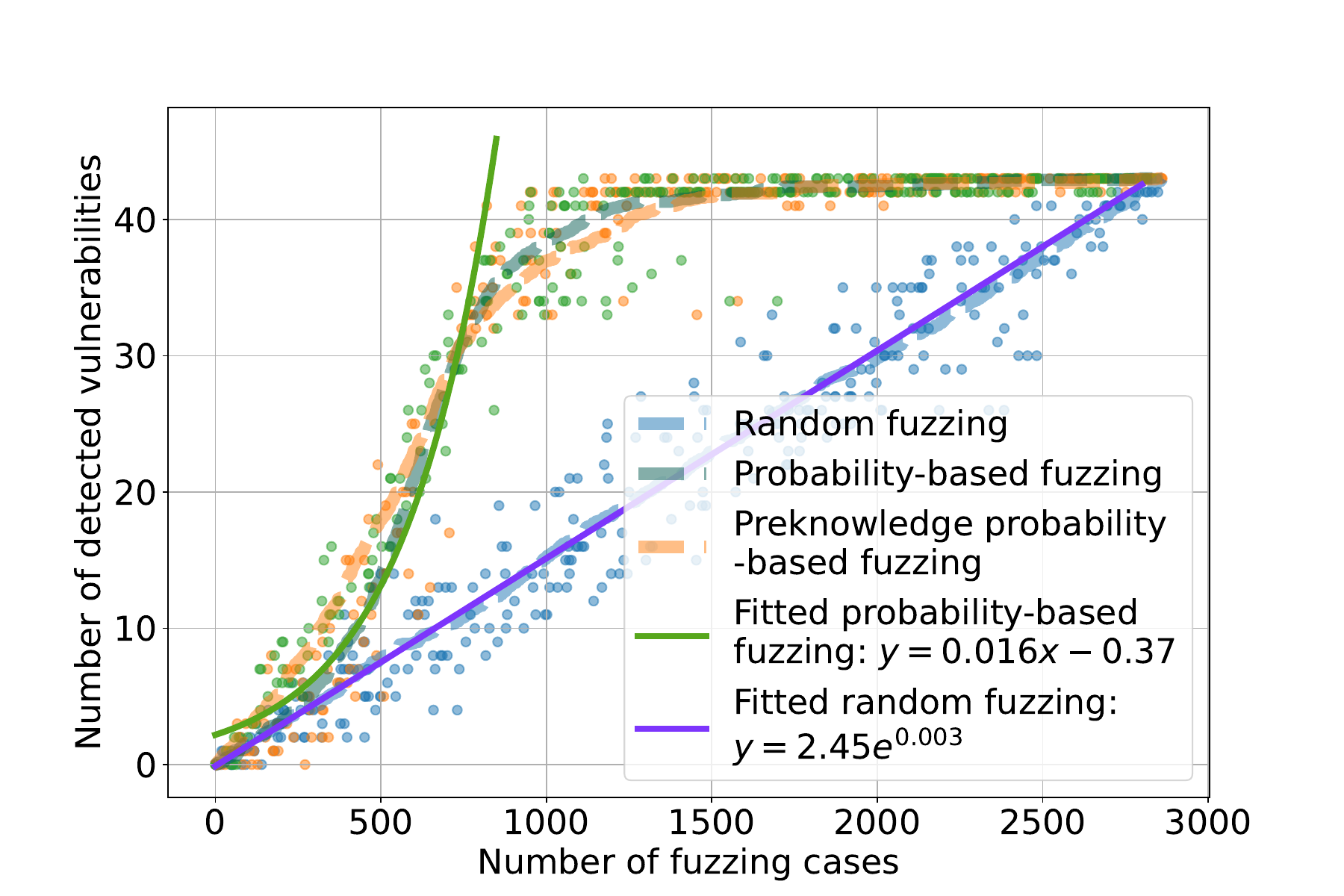}
    \caption{Comparison of Benchmark random-based fuzzing and probability-based fuzzing.\cite{yang2023systematic}}
    \label{fig:Comparison_command}
    \vspace{-15pt}
\end{figure}

In addition to vulnerability detection at the bit-level of a command using fuzzing, it is also necessary to verify formal attack traces using command-level fuzzing. Unlike bit-level fuzzing, no representative case can cover all out-of-rule cases, which means there are an unlimited number of cases in command-level fuzzing. To efficiently locate command-level vulnerabilities, we proposed a probability-based command-level fuzzing framework in our previous work~\cite{yang2023systematic}. Based on formal assumptions of \ac{RRC} and User Identity Disclosure attack, we fixed the C-RNTI and ISMI on the srsRAN platform to simulate the disclosure of user identity. This reduced the number of fuzzing cases to 3,080. Furthermore, based on the identity disclosure assumption, we collected all different commands on downlink channels and fuzzed all possible permutations. From Fig.~\ref{fig:Comparison_command}, we can conclude that our proposed probability-based framework requires only $36.5\%$ of the fuzzing cases numbers using a random fuzzing strategy. The number of cases needed for different percentages of detected vulnerabilities is shown. In comparison to the conventional linear growth computation-consuming random fuzzing strategy, we developed probability-based fuzzing approach demonstrates significantly improved performance\cite{yang2023systematic}. The incorporation of prior knowledge further enhances the effectiveness of our method, leading to even greater efficiency gains. Theoretically, our proposed approaches have the potential to complete millions of command-level fuzzing iterations within a modest scope of five thousand test cases. This significant reduction in the number of required test cases underscores the efficiency and effectiveness of our methodology.

\section{Conclusion and Future Work}

Motivated by the limitations of state-of-the-art vulnerability detection methods, which include highly computational complex and labor-intensive formal and fuzz testing approaches, in this paper, we present a first-of-its-kind formal guided fuzz testing approach for an efficient and systematic 5G vulnerability detection. In particular, in the proposed approach, formal verification is implemented to detect attack traces in 5G protocols, which are then utilized to guide subsequent fuzz testing.

We demonstrate the detection of $4$ attack models and $61$ vulnerabilities in 5G NSA authentication and authorization procedure. We present the generality and stability of applying the formal guided fuzz testing framework to provide assurance in other protocols in 5G and Future G releases. The four attack model categories, which include modification of \ac{RRC} connection, \ac{DoS} or device disconnection using Authentication Request, exposure of $K_{NASenc}$ and $K_{NASint}$, and exposure of $K_{RRCenc}$, $K_{RRCint}$, and $K_{UPenc}$, are extrapolated from the attack traces inferred through formal verification. The detected vulnerabilities by guided fuzz testing further identify the risks and impacts in each of the four attack models and are verified via real-life experiments using srsRAN. The detected attack model discovery and vulnerability detection include the exploits discussed in existing research and new findings that have never been revealed. Our approach connects the strengths and coverage of formal and fuzzing methods to efficiently detect vulnerabilities across protocol logic and implementation stacks hierarchically and interactively. To close the loop, we incorporate feedback from the detected attack models and vulnerabilities to fortify systems designs and enhance system resilience. This innovative approach enables the auto-discovery of vulnerabilities and unintended emergent behaviors from the communications specifications to implementation stacks.


Furthermore, in addressing the computation complexity, we assess the complexity of our approach with conventional fuzz testing results and the state of art approaches. Conventional fuzz testing would necessitate a staggering $9\times10^{77}$ fuzzing cases. Latest researches~\cite{Salazar20215Greplay:Injection,he2022intelligent} reveal that identifier-specific rule-based fuzz testing would require a lesser, yet substantial, $6\times10^{38}$ fuzzing cases. In contrast, our system uncovers $8$ vulnerabilities within a mere $42$ representative fuzz testing cases under the guidance of formal verification, thereby demonstrating its bit-level vulnerability detection proficiency. In the realm of command-level fuzzing, out-of-rule cases are infinite. However, under the formal assumption of RRC and User Identity Disclosure attack, our framework reduces the number of fuzzing cases to a manageable $3080$, which is further curtailed to $1027$ through probability-based fuzzing strategy, showcasing the framework's superior efficiency.

In the future, we will transfer the framework into an automatic multi-dimensions vulnerability detection system with reinforcement loop feedback. The new model will consider a wider variety of data to enable multi-dimensional input and analysis, like log files and the state of the cache. In addition to 5G specifications, we will expand the verification and vulnerability detection to various specifications and implementations, including IoT and other areas.

\section*{Acknowledgment}
The views and conclusions contained herein are those of the authors and should not be interpreted as necessarily representing the official policies or endorsements, either expressed or implied, of DARPA or the U.S. Government.

\bibliographystyle{IEEEtran}
\bibliography{reference,references_ying}

\begin{thebibliography}{10}
\providecommand{\url}[1]{#1}
\csname url@samestyle\endcsname
\providecommand{\newblock}{\relax}
\providecommand{\bibinfo}[2]{#2}
\providecommand{\BIBentrySTDinterwordspacing}{\spaceskip=0pt\relax}
\providecommand{\BIBentryALTinterwordstretchfactor}{4}
\providecommand{\BIBentryALTinterwordspacing}{\spaceskip=\fontdimen2\font plus
\BIBentryALTinterwordstretchfactor\fontdimen3\font minus
  \fontdimen4\font\relax}
\providecommand{\BIBforeignlanguage}[2]{{%
\expandafter\ifx\csname l@#1\endcsname\relax
\typeout{** WARNING: IEEEtran.bst: No hyphenation pattern has been}%
\typeout{** loaded for the language `#1'. Using the pattern for}%
\typeout{** the default language instead.}%
\else
\language=\csname l@#1\endcsname
\fi
#2}}
\providecommand{\BIBdecl}{\relax}
\BIBdecl

\bibitem{Alcaraz-Calero2018Leading5G-PPP}
J.~Alcaraz-Calero, I.~P. Belikaidis, C.~J.~B. Cano, P.~Bisson, D.~Bourse,
  M.~Bredel, D.~Camps-Mur, T.~Chen, X.~Costa-Perez, P.~Demestichas, M.~Doll,
  S.~E. Elayoubi, A.~Georgakopoulos, A.~M{\"{a}}mmel{\"{a}}, H.~P. Mayer,
  M.~Payaro, B.~Sayadi, M.~S. Siddiqui, M.~Tercero, and Q.~Wang, ``{Leading
  innovations towards 5G: Europe's perspective in 5G Infrastructure
  Public-Private Partnership (5G-PPP)},'' in \emph{IEEE International Symposium
  on Personal, Indoor and Mobile Radio Communications, PIMRC}, vol.
  2017-October, 2018.

\bibitem{shatnawi2022digital}
M.~Shatnawi, H.~Altaleb, and R.~Zolt{\'a}n, ``The digital revolution with nesas
  assessment and evaluation,'' in \emph{2022 IEEE 10th Jubilee International
  Conference on Computational Cybernetics and Cyber-Medical Systems
  (ICCC)}.\hskip 1em plus 0.5em minus 0.4em\relax IEEE, 2022, pp.
  000\,099--000\,104.

\bibitem{Hussain20195Greasoner:Protocol}
S.~R. Hussain, M.~Echeverria, I.~Karim, O.~Chowdhury, and E.~Bertino,
  ``{5Greasoner: A property-directed security and privacy analysis framework
  for 5G cellular network protocol},'' in \emph{Proceedings of the ACM
  Conference on Computer and Communications Security}.\hskip 1em plus 0.5em
  minus 0.4em\relax Association for Computing Machinery, 11 2019, pp. 669--684.

\bibitem{klees2018evaluating}
G.~Klees, A.~Ruef, B.~Cooper, S.~Wei, and M.~Hicks, ``Evaluating fuzz
  testing,'' in \emph{Proceedings of the 2018 ACM SIGSAC conference on computer
  and communications security}, 2018, pp. 2123--2138.

\bibitem{souri2019state}
A.~Souri and M.~Norouzi, ``A state-of-the-art survey on formal verification of
  the internet of things applications,'' \emph{Journal of Service Science
  Research}, vol.~11, no.~1, pp. 47--67, 2019.

\bibitem{beaman2022fuzzing}
C.~Beaman, M.~Redbourne, J.~D. Mummery, and S.~Hakak, ``Fuzzing vulnerability
  discovery techniques: survey, challenges and future directions,''
  \emph{Computers \& Security}, p. 102813, 2022.

\bibitem{Wang2021AI-PoweredOptimization}
Y.~Wang, A.~Gorski, and L.~A. DaSilva, ``{AI-Powered Real-Time Channel
  Awareness and 5G NR Radio Access Network Scheduling Optimization},'' 2021.

\bibitem{Wang2022AnonymousShetty}
Y.~Wang, S.~Jere, S.~Banerjee, L.~Liu, V.~Modeling, and S.~Dayekh, ``{Anonymous
  Jamming Detection in 5G with Bayesian Network Model Based Inference Analysis
  Sachin Shetty},'' in \emph{IEEE International Conference on High Performance
  Switching and Routing 6–8 June 2022 // Virtual Conference}, 2022.

\bibitem{Meier2013TheProtocols}
S.~Meier, B.~Schmidt, C.~Cremers, and D.~Basin, ``{The TAMARIN prover for the
  symbolic analysis of security protocols},'' in \emph{Lecture Notes in
  Computer Science (including subseries Lecture Notes in Artificial
  Intelligence and Lecture Notes in Bioinformatics)}, vol. 8044 LNCS, 2013.

\bibitem{Cremers2019Component-BasedConfusion}
C.~Cremers and M.~Dehnel-Wild, ``{Component-Based Formal Analysis of 5G-AKA:
  Channel Assumptions and Session Confusion}.''\hskip 1em plus 0.5em minus
  0.4em\relax Internet Society, 3 2019.

\bibitem{peltonen2021comprehensive}
A.~Peltonen, R.~Sasse, and D.~Basin, ``A comprehensive formal analysis of 5g
  handover,'' in \emph{Proceedings of the 14th A CM Conference on Security and
  Privacy in Wireless and Mobile Networks}, 2021, pp. 1--12.

\bibitem{Labib2017EnhancingProcess}
M.~Labib, V.~Marojevic, J.~H. Reed, and A.~I. Zaghloul, ``{Enhancing the
  Robustness of LTE Systems: Analysis and Evolution of the Cell Selection
  Process},'' \emph{IEEE Communications Magazine}, vol.~55, no.~2, 2017.

\bibitem{Rupprecht2019BreakingTwo}
D.~Rupprecht, K.~Kohls, T.~Holz, and C.~Popper, ``{Breaking LTE on Layer
  Two},'' in \emph{Proceedings - IEEE Symposium on Security and Privacy}, vol.
  2019-May, 2019.

\bibitem{Shaik2017PracticalSystems}
A.~Shaik, R.~Borgaonkar, N.~Asokan, V.~Niemi, and J.-P. Seifert, ``{Practical
  Attacks Against Privacy and Availability in 4G/LTE Mobile Communication
  Systems},'' 2017.

\bibitem{Basin2018AAuthentication}
D.~Basin, S.~Radomirovic, J.~Dreier, R.~Sasse, L.~Hirschi, and V.~Stettler,
  ``{A formal analysis of 5g authentication},'' in \emph{Proceedings of the ACM
  Conference on Computer and Communications Security}.\hskip 1em plus 0.5em
  minus 0.4em\relax Association for Computing Machinery, 10 2018, pp.
  1383--1396.

\bibitem{Klees2018EvaluatingTesting}
G.~Klees, A.~Ruef, B.~Cooper, S.~Wei, and M.~Hicks, ``{Evaluating fuzz
  testing},'' in \emph{Proceedings of the ACM Conference on Computer and
  Communications Security}, 2018.

\bibitem{wang2022automated}
H.~Wang, B.~Cui, W.~Yang, J.~Cui, L.~Su, and L.~Sun, ``An automated
  vulnerability detection method for the 5g rrc protocol based on fuzzing,'' in
  \emph{2022 4th International Conference on Advances in Computer Technology,
  Information Science and Communications (CTISC)}.\hskip 1em plus 0.5em minus
  0.4em\relax IEEE, 2022, pp. 1--7.

\bibitem{he2022intelligent}
F.~He, W.~Yang, B.~Cui, and J.~Cui, ``Intelligent fuzzing algorithm for 5g nas
  protocol based on predefined rules,'' in \emph{2022 International Conference
  on Computer Communications and Networks (ICCCN)}.\hskip 1em plus 0.5em minus
  0.4em\relax IEEE, 2022, pp. 1--7.

\bibitem{moukahal2021vulnerability}
L.~J. Moukahal, M.~Zulkernine, and M.~Soukup, ``Vulnerability-oriented fuzz
  testing for connected autonomous vehicle systems,'' \emph{IEEE Transactions
  on Reliability}, vol.~70, no.~4, pp. 1422--1437, 2021.

\bibitem{han2012mutation}
X.~Han, Q.~Wen, and Z.~Zhang, ``A mutation-based fuzz testing approach for
  network protocol vulnerability detection,'' in \emph{Proceedings of 2012 2nd
  International conference on computer science and network technology}.\hskip
  1em plus 0.5em minus 0.4em\relax IEEE, 2012, pp. 1018--1022.

\bibitem{Salazar20215Greplay:Injection}
Z.~Salazar, H.~N. Nguyen, W.~Mallouli, A.~R. Cavalli, and E.~M. Montes De~Oca,
  ``{5Greplay: A 5G Network Traffic Fuzzer - Application to Attack
  Injection},'' in \emph{ACM International Conference Proceeding Series}.\hskip
  1em plus 0.5em minus 0.4em\relax Association for Computing Machinery, 8 2021.

\bibitem{sheikhi2022coverage}
S.~Sheikhi, E.~Kim, P.~S. Duggirala, and S.~Bak, ``Coverage-guided fuzz testing
  for cyber-physical systems,'' in \emph{2022 ACM/IEEE 13th International
  Conference on Cyber-Physical Systems (ICCPS)}.\hskip 1em plus 0.5em minus
  0.4em\relax IEEE, 2022, pp. 24--33.

\bibitem{ammann2023dy}
M.~Ammann, L.~Hirschi, and S.~Kremer, ``Dy fuzzing: Formal dolev-yao models
  meet protocol fuzz testing,'' \emph{Cryptology ePrint Archive}, 2023.

\bibitem{ma2017semi}
R.~Ma, S.~Ren, K.~Ma, C.~Hu, and J.~Xue, ``Semi-valid fuzz testing case
  generation for stateful network protocol,'' \emph{Tsinghua Science and
  Technology}, vol.~22, no.~5, pp. 458--468, 2017.

\bibitem{bratus2008lzfuzz}
S.~Bratus, A.~Hansen, and A.~Shubina, ``Lzfuzz: a fast compression-based fuzzer
  for poorly documented protocols,'' 2008.

\bibitem{osborne2021leveraging}
N.~Osborne and C.~Pascutto, ``Leveraging formal specifications to generate
  fuzzing suites,'' in \emph{OCaml Users and Developers Workshop, co-located
  with the 26th ACM SIGPLAN International Conference on Functional
  Programming}, 2021.

\bibitem{yang2023systematic}
J.~Yang, Y.~Wang, Y.~Pan, and T.~X. Tran, ``Systematic meets unintended: Prior
  knowledge adaptive 5g vulnerability detection via multi-fuzzing,''
  \emph{arXiv preprint arXiv:2305.08039}, 2023.

\bibitem{O-RANAlliance2018O-RAN:RAN}
{O-RAN Alliance}, ``{O-RAN: Towards an Open and Smart RAN},'' \emph{O-RAN
  Alliance}, no. October, 2018.

\bibitem{SoftwareRadioSystems2021SrsRANSRS}
{Software Radio Systems}, ``{srsRAN is a 4G/5G software radio suite developed
  by SRS},'' 2021.

\bibitem{Wang2021DevelopmentResearch}
Y.~Wang, A.~Gorski, and A.~da~Silva, ``{Development of a Data-Driven Mobile 5G
  Testbed: Platform for Experimental Research},'' in \emph{IEEE International
  Mediterranean Conference on Communications and Networking}, 2021.

\bibitem{wang2012dynamic}
Y.~Wang and C.~W. Bostian, ``Dynamic cellular cognitive system,'' Jan.~10 2012,
  uS Patent 8,094,610.

\bibitem{JingdaYang20235GListen-and-Learn}
{Jingda Yang}, {Ying Wang}, {Tuyen X. Tran}, and {Yanjun Pan}, ``{5G RRC
  Protocol and Stack Vulnerabilities Detection via Listen-and-Learn},'' in
  \emph{IEEE Consumer Communications {\&} Networking Conference}, 2023.

\bibitem{Dauphinais2023AutomatedSystems}
D.~Dauphinais, M.~Zylka, H.~Spahic, F.~k. Shai, J.~Yang, I.~Cruz, J.~Gibson,
  and Y.~Wang, ``{Automated Vulnerability Testing and Detection Digital Twin
  Framework for 5G Systems},'' in \emph{9th IEEE International Conference on
  Network Softwarization}, Madrid, 6 2023.

\bibitem{Yang2023SystematicImplementations}
J.~Yang, Y.~Wang, Y.~Pan, and T.~X. Tran, ``{Systematic and Scalable
  Vulnerability Detection for 5G Specifications and Implementations},''
  \emph{IEEE JOURNAL ON SELECTED AREAS IN COMMUNICATIONS(Under Review)}, 2023.

\bibitem{gomez2016srslte}
I.~Gomez-Miguelez, A.~Garcia-Saavedra, P.~D. Sutton, P.~Serrano, C.~Cano, and
  D.~J. Leith, ``srslte: An open-source platform for lte evolution and
  experimentation,'' in \emph{Proceedings of the Tenth ACM International
  Workshop on Wireless Network Testbeds, Experimental Evaluation, and
  Characterization}, 2016, pp. 25--32.

\bibitem{6235098}
J.~B. Bou~Abdo, H.~Chaouchi, and M.~Aoude, ``Ensured confidentiality
  authentication and key agreement protocol for eps,'' in \emph{2012 Symposium
  on Broadband Networks and Fast Internet (RELABIRA)}, 2012, pp. 73--77.

\bibitem{3gpp.29.118}
\BIBentryALTinterwordspacing
3GPP, ``{Universal Mobile Telecommunications System (UMTS); LTE; Mobility
  Management Entity (MME) Visitor Location Register (VLR) SGs interface
  specification},'' {3rd Generation Partnership Project (3GPP)}, Technical
  Specification (TS) 29.118, 01 2015, version 8.5.0. [Online]. Available:
  \url{https://portal.3gpp.org/desktopmodules/Specifications/SpecificationDetails.aspx?specificationId=1601}
\BIBentrySTDinterwordspacing

\bibitem{khan2018defeating}
M.~Khan, P.~Ginzboorg, K.~J{\"a}rvinen, and V.~Niemi, ``Defeating the downgrade
  attack on identity privacy in 5g,'' in \emph{International Conference on
  Research in Security Standardisation}.\hskip 1em plus 0.5em minus 0.4em\relax
  Springer, 2018, pp. 95--119.

\bibitem{tsay2012vulnerability}
J.-K. Tsay and S.~F. Mj{\o}lsnes, ``A vulnerability in the umts and lte
  authentication and key agreement protocols,'' in \emph{Computer Network
  Security: 6th International Conference on Mathematical Methods, Models and
  Architectures for Computer Network Security, MMM-ACNS 2012, St. Petersburg,
  Russia, October 17-19, 2012. Proceedings 6}.\hskip 1em plus 0.5em minus
  0.4em\relax Springer, 2012, pp. 65--76.

\bibitem{raza2018exposing}
M.~T. Raza, F.~M. Anwar, and S.~Lu, ``Exposing lte security weaknesses at
  protocol inter-layer, and inter-radio interactions,'' in \emph{Security and
  Privacy in Communication Networks: 13th International Conference, SecureComm
  2017, Niagara Falls, ON, Canada, October 22--25, 2017, Proceedings 13}.\hskip
  1em plus 0.5em minus 0.4em\relax Springer, 2018, pp. 312--338.

\end{thebibliography}

\begin{IEEEbiography}[{\includegraphics[width=1in,height=1.25in,clip,keepaspectratio]{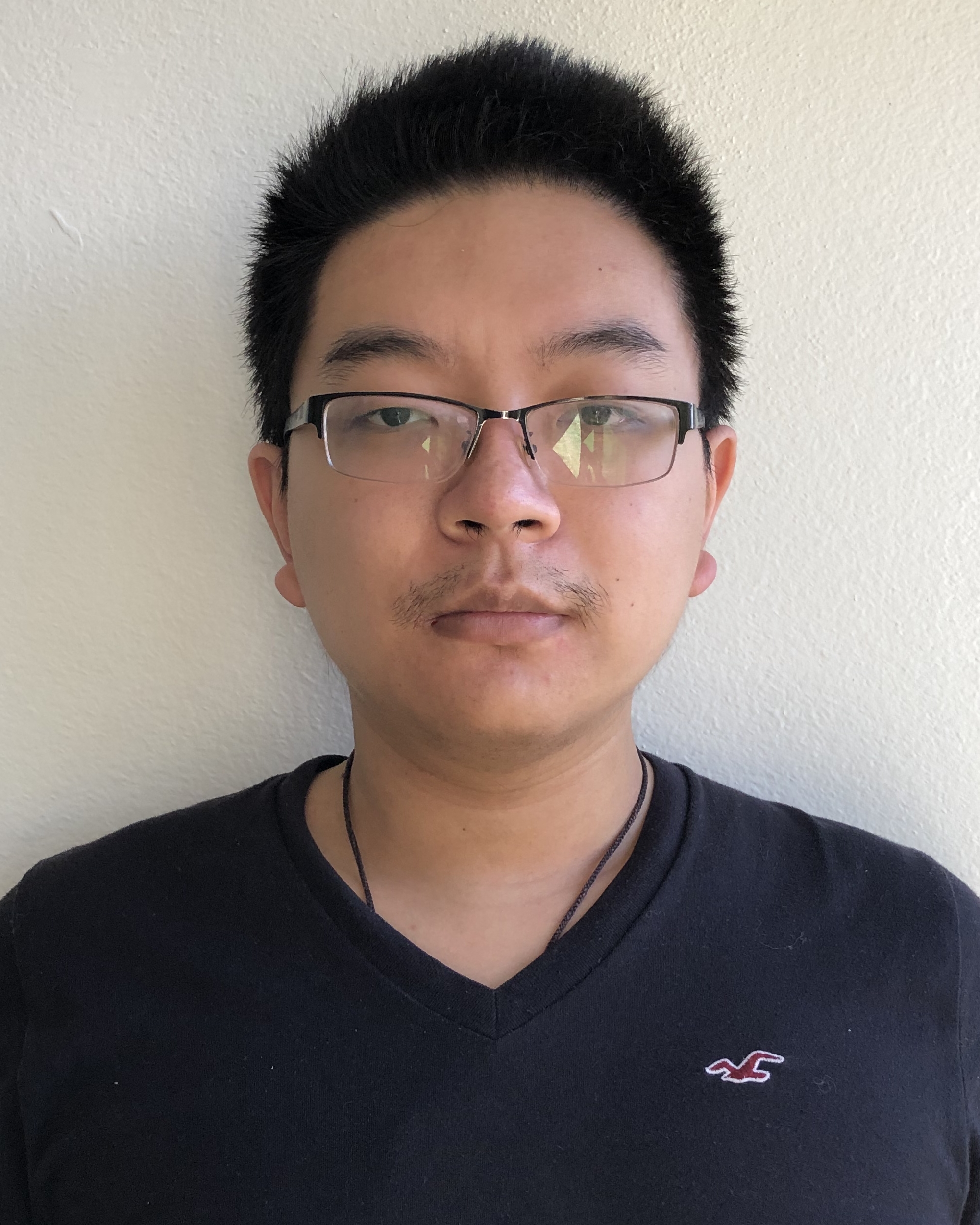}}]{Jingda Yang} (Graduate Student Member, IEEE) received the B.E. degree in software engineering from Shandong University and the M.Sc. degree in computer science from The George Washington University. He is currently a Ph.D. student in the School of System and Enterprises at Stevens Institute of Technology.  His research interests are formal verification and vulnerability detection of wireless protocol in 5G.
\end{IEEEbiography}

\begin{IEEEbiography}[{\includegraphics[width=1.0in,height=1.25in, clip,keepaspectratio]{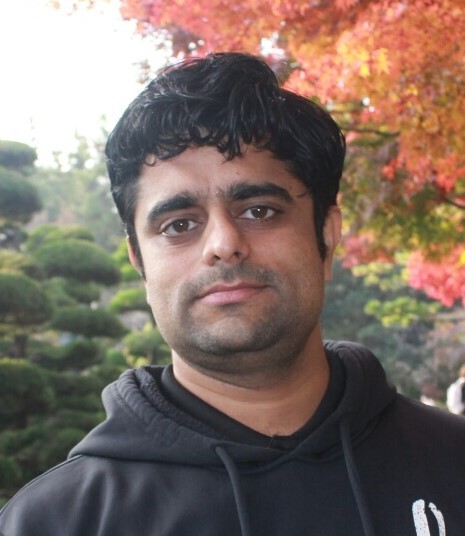}}]{Sudhanshu Arya} (Member, IEEE) is a Research Fellow in the School of System and Enterprises at Stevens Institute of Technology, NJ, USA. He received his M.Tech. degree in communications and networks from the National Institute of Technology, Rourkela, India, in 2017, and the Ph.D. degree from Pukyong National University, Busan, South Korea, in 2022. He worked as a Research Fellow with the Department of Artificial Intelligence Convergence, at Pukyong National University. His research interests include wireless communications and digital signal processing, with a focus on free-space optical communications, optical scattering communications, optical spectrum sensing, computational game theory, and artificial intelligence. He received the Best Paper Award in ICGHIT 2018 and the Early Career Researcher Award from the Pukyong National University in 2020.
\end{IEEEbiography}

\begin{IEEEbiography}[{\includegraphics[width=1in,height=1.25in,clip,keepaspectratio]{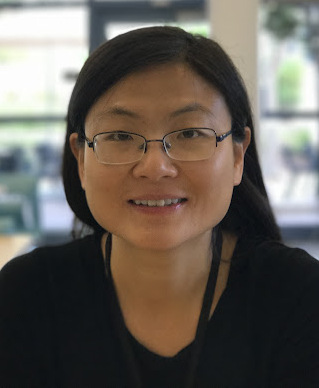}}]{Ying Wang} (Member, IEEE) received the B.E. degree in information engineering at Beijing University of Posts and Telecommunications, M.S. degree in electrical engineering from University of Cincinnati and the Ph.D. degree in electrical engineering from Virginia Polytechnic Institute and State University. She is an associate professor in the School of System and Enterprises at Stevens Institute of Technology. Her research areas include cybersecurity, wireless AI, edge computing, health informatics, and software engineering. 
\end{IEEEbiography}

\end{document}